\documentclass[12pt,preprint]{aastex}

\shorttitle{Cold Gas in Galactic Winds}
\shortauthors{Fujita et al.}

\begin{document}
\title{The Origin and Kinematics 
of Cold Gas in Galactic Winds: Insight from Numerical Simulations}

\author{Akimi Fujita\altaffilmark{1,2}, Crystal
L. Martin\altaffilmark{1,6,7}, Mordecai-Mark Mac
Low\altaffilmark{3,2,4}, Kimberly C. B. New\altaffilmark{5}, and
Robert Weaver\altaffilmark{5}}

\altaffiltext{1}{Department of Physics, University of California, Santa
  Barbara, CA  93106; cmartin@physics.ucsb.edu} 
\altaffiltext{2}{Max-Planck-Institut f\"ur Astronomie, 69117 Heidelberg,
  Germany} 
\altaffiltext{3}{Department of Astrophysics, American Museum of Natural
  History, New York, NY, 10024; mordecai@amnh.org}
\altaffiltext{4}{Institut f\"ur Theoretische Astrophysik, Zentrum f\"ur
  Astronomie der Universit\"at Heidelberg, 69120 Heidelberg, Germany}
\altaffiltext{5}{Los Alamos National Laboratory, Los Alamos, NM, 87545}
\vspace{0.2in}
\altaffiltext{6}{Packard Fellow}
\altaffiltext{7}{Alfred P. Sloan Foundation Fellow}

\begin{abstract}
We study the origin of Na~{\sc i} absorbing gas in ultraluminous
infrared galaxies motivated by the recent observations by Martin of
extremely superthermal linewidths in this cool gas.  We model the
effects of repeated supernova explosions driving supershells in the
central regions of molecular disks with $M_d=10^{10} M_{\odot}$,
using cylindrically symmetric gas dynamical simulations run with
ZEUS-3D.  The shocked swept-up shells quickly cool and fragment by
Rayleigh-Taylor instability as they accelerate out of the dense,
stratified disks.  The numerical resolution of the cooling and
compression at the shock fronts determines the peak shell density,
and so the speed of Rayleigh-Taylor fragmentation.  We identify
cooled shells and shell fragments as Na~{\sc i} absorbing gas and
study its kinematics along various sightlines across the grid.  We
find that simulations with a numerical resolution of $\le 0.2$ pc
produce multiple Rayleigh-Taylor fragmented shells in a given line
of sight that appear to explain the observed kinematics.  We suggest
that the observed wide Na~{\sc i} absorption lines, $\langle v
\rangle =320\pm120\mbox{ km s}^{-1}$ are produced by these multiple
fragmented shells traveling at different velocities.  We also
suggest that some shell fragments can be accelerated above the
observed average terminal velocity of $750\mbox{ km s}^{-1}$ by the
same energy-driven wind with an instantaneous starburst of
$\sim10^9~M_{\odot}$.  The mass carried by these fragments is only a
small fraction of the total shell mass, while the bulk of mass is
traveling with velocities consistent with the observed average shell
velocity $330\pm100\mbox{ km s}^{-1}$.  Our results show that an
energy-driven bubble causing Rayleigh-Taylor instabilities can
explain the kinematics of cool gas seen in the Na~{\sc i}
observations without invoking additional physics relying primarily
on momentum conservation, such as entrainment of gas by
Kelvin-Helmholtz instabilities, ram pressure driving of cold clouds
by a hot wind, or radiation pressure acting on dust.
\end{abstract}

\keywords{hydrodynamics, supernovae: general, ISM: bubbles, galaxies:
  starburst, ISM: jets and outflows, ISM: kinematics and dynamics}

\section{Introduction}

Nearly all starburst galaxies, regardless of mass, appear to drive
large-scale gaseous outflows, or galactic winds (Heckman et al.\ 1990;
Martin 1999).  Measurements demonstrate that these metal-enriched
winds transport interstellar gas and supernova ejecta into galactic
halos (Martin, et al.\ 2002). These winds are thought to influence the
thermal and chemical evolution of the intergalactic medium and hence
the formation of galaxies as well as their evolution.

From radio to X-ray frequencies, observations of starburst galaxies
reveal outflowing gas over a very broad temperature range (Martin et
al.\ 2002). However, all observed emission is relatively near the
galaxy, within a projected separation of about 10~kpc, due to the
radial density gradient of the wind and density-squared dependence of
emission processes. Absorption-line measurements are more sensitive to
extended, low-density gas. The number of detections of blue-shifted
(i.e. outflowing) interstellar absorption lines in starburst galaxy
spectra has grown by a large factor in recent years (Heckman et al.\
2000; Rupke et al.\ 2002; Schwartz \& Martin 2004; Martin 2005). The
shortcoming of absorption line measurements is that they do not
uniquely determine the distance between the galaxy and the absorbing
material.

Numerical simulations of galactic winds can provide needed insight
into where the absorption originates.  Using simulations to interpret
observations, and observations to constrain simulations, is probably
the only way to really understand these complex outflows dynamically.
Modeling the early evolution of a galactic wind as it blows out of its
disk requires a numerical, rather than analytic, approach due to the
importance of nonlinear hydrodynamic and thermal instabilities.

Supershells evolve with roughly spherical geometry until they grow to
scales of the disk gas scale-height (Tomisaka \& Ikeuchi 1986, 1988;
Mac Low \& McCray 1988; Tenorio-Tagle \& Bodenheimer 1988; de Young \&
Heckman 1994).  The acceleration of the shell into the galactic halo
causes it to fragment via Rayleigh-Taylor (R-T) instabilities (Mac
Low et al.\ 1989).  The hot, low-density bubble interior
radiates inefficiently.  The wind can sweep up new shells of ambient
gas, that in turn fragment by R-T instability, leaving a broad region
containing fragments of fast-moving cool gas.

The swept up shell is driven by the thermal pressure of the interior
$P = \rho c_s^2$, where $c_s$ is the interior sound speed. After
blowout, the hot gas expands freely through the fragmented shell,
producing a supersonic, energy-driven wind with velocity $v_w$.
Although entrained shell fragments can still be accelerated by the ram
pressure of the wind $P_{ram} = \rho v_w^2$, this appears to be a
minor contribution to their total kinetic energy.  This can be seen by
comparing the velocity of a bubble expanding into a uniform medium at
a radius of one scale height to the final shell fragment velocities,
as reported, for example, by Mac Low et al.\ (1989). These are the
same, to within a factor of two.

Properties of the cool gas in starburst winds have been estimated from
observations of interstellar Na~{\sc i} lines in starburst galaxy spectra.
Estimates of the total mass of cold gas in these outflows have large 
uncertainties at present due to line saturation at low resolution as
well as corrections for ionization state and dust depletion. Nonetheless,
it has been emphasized that the momentum of the cool flows appear to be
somewhat less than the amount available from either supernova ejecta or the 
radiation field, at least for the most luminous starbursts (Rupke et
al.\ 2005; Martin 2006).  The same approximations, however, also yield kinetic
energies for the cool outflow that are only a few percent (up to a 
few tens of percent) of the supernova energy.  The same flows could also
be driven by energy-conserving bubbles, with only a small fraction of the 
total energy in the bubble going to accelerating the cold gas.

The standard scenario used for interpreting starburst wind absorption
is based on the simulation shown in Figure~11 of Heckman et al.\
(2000), which suggests that dense clouds are advected into the wind at
the interface between the low-latitude disk and the wind, by
Kelvin-Helmholtz instability.  With a grid resolution of 4.9 pc,
however, the clumps of dense gas are not fully resolved in that
simulation, leaving artificially large clumps that completely stop
fragmenting below $\sim6$ zone size.  We will show below that the R-T
instability is suppressed in secondary shells at that resolution, as
well, substantially changing the distribution of cool gas.  

Recently, Cooper et al.\ (2008) performed three-dimensional (3D)
simulations of starburst blowout through a galactic disk with a
fractal density distribution.  They injected energy at a rate
proportional to local density, rather than identifying supernova sites
and following the explicit evolution of their remnants.  This leads to
higher than physical radiative losses, so their results represent a
lower limit to the effects of a starburst.  They found that H$\alpha$
emitting gas comes from the gas dynamical stripping and fragmentation
of existing interstellar clouds.  This gas can reasonably also be
identified as a potential source of absorption, although they did not
address the question explicitly, nor extend the simulation to times
long enough to directly model the absorption.

We instead identify the location of the absorbing gas in fragmenting
shells of swept-up interstellar gas, using high-resolution
two-dimensional (2D) simulations with resolution as small as 0.1~pc.
Although in our models we compute only up to the time of blowout
because of the small region covered by our computational grid, we use
the ballistic approximation (Zahnle \& Mac Low 1995; Fujita et al.\
2004) to show that gravitational deceleration does not act strongly on
the shells and fragments during the starburst duration (10--40 Myr) if
their velocities exceed $50-200 \mbox{ km s}^{-1}$ at blowout.  This
analysis is based, however, on an assumption that the bulk of their
mass remains unablated by the wind blowing past them.  Understanding
the full history of shell fragments and clumps will require
substantial further work.

We address the origin and kinematics of the cold wind as measured in
the Na~{\sc i} $\lambda\lambda$5890, 96 absorption lines.  The
observations pose three major questions. First, why do the absorption
line widths tend to greatly exceed the thermal velocity dispersion of
warm neutral gas?  The average full width at half maximum (FWHM) of
the dynamic component is $320\pm120\mbox{ km s}^{-1}$ in ultraluminous
infrared galaxies (ULIRGs), while the line widths range from 150 to
$600 \mbox{ km s}^{-1}$ in luminous infrared galaxies.  Second, why do
the terminal velocities of the cold gas approach the escape velocities
from the starburst galaxies (Martin 2005)?  Third, what do the maximum
and mean velocities measured in the line profiles really represent
physically?

We use our models to pursue
five investigations. First, we investigate
how the absorption properties change with viewing angle. We
specifically test whether multiple R-T fragmented shells along a line
of sight can reproduce the broad line width seen in Na~{\sc i}
absorption lines.  Second, we vary the 
   numerical 
resolution to demonstrate how increased resolution of radiative
cooling behind the shocks, and so of shell fragmentation, affect the
results.

Third, we make a more general parameter study addressing variations in
the properties of the outflowing cold gas with starburst luminosity,
the size of the starburst region, and gas surface density.  Fourth, we
can obtain insight into the complicated dynamics of multiphase
outflows, particularly their dependence on the mass-loading of the
wind.  We investigate mass-loading rates between
$\sim1.7-120~M_{\odot}\mbox{ yr}^{-1}$ and vary the mechanical
luminosity of the starburst between $10^{41} - 10^{43}$~erg~s$^{-1}$
to see what velocities are reached by the swept-up shells and their
fragments.  The observed X-ray temperatures vary little with starburst
luminosity $T \sim 10^7 K$, so the terminal wind velocities should
vary little with luminosity (above some critical value required for
blowout).

Finally, Heckman et al.\ (2000) argued that both Na~{\sc i} absorbing
and H$\alpha$ emitting gas can not originate in the swept-up shells
because of the lack of strong correlation between the widths of
Na~{\sc i} absorption lines and H$\alpha$ emission lines.  For
example, the outflow sources with very broad (400--600 km s$^{-1}$)
Na~{\sc i} absorption lines have H$\alpha$ emission-line widths
ranging from 145 to 1500 km s$^{-1}$.  Although a full nebular
emission calculation is well beyond the scope of this paper, we do
discuss where the ionization front might reside for various ionizing
photon luminosities.  We study the kinematics of the low-ionization
Na~{\sc i} absorbing gas and photoionized H$\alpha$ emitting gas by
separating them crudely, using the photoionization code of Abel et
al.\ (1999).

The acceleration of shell fragments is sensitive to how well shell
fragmentation is resolved.  Applying adaptive mesh refinement (AMR)
techniques to this problem can maintain high resolution in the shocked
shells and clouds. This paper is the first step toward such an
improved simulation.  We compare the blowout problem run on a fixed
grid to a similar problem run with an adaptive grid, focusing on the
comparison to measured properties of cold gas in galactic winds.

In this paper, we describe our disk and star formation models in \S~2
and our numerical method in \S~3.  We give the results of our
parameter studies in \S~4 and discuss comparisons with observations in
\S~5, followed by conclusions in \S~6.  In an Appendix, we show the
results of test simulations of blowout in a dwarf galaxy by ZEUS-3D
(Stone \& Norman 1992a; Clarke 1994) and SAGE (SAIC's Adaptive Grid,
Eulerian hydrocode; Kerbyson et al.\ 2001; Gittings et al.\ 2008).

\section{Disk and Star Formation Models}

The parameters of our starburst model are based on the properties of
ULIRGs to facilitate comparison with Martin (2005, 2006).  We use
hydrodynamic simulations to model the effects of multiple supernova
explosions in the central 200~pc $\times 100$~pc region of the
molecular disk of a ULIRG.  Our model is an extension of the blowout
model in dwarf galaxies described by Mac Low \& Ferrara (1999) and
Fujita et al.\ (2003).  Our fiducial numerical resolution is 0.2 pc,
sufficient to resolve cooling behind the shocks, and so the
fragmentation of the swept-up shells by R-T instability as well as
possible within a reasonable computational time. 
  To study the effects of numerical resolution, we use models with
  resolution ranging from 0.1 to 0.8 pc.

\subsection{Disk}
ULIRGs are starburst galaxies with infrared luminosity $>10^{12}
L_{\odot}$, and are usually found in major mergers and interacting
galaxies (Sanders et al.\ 1988).  They are believed to go through
starburst phases twice, when the gas in a galaxy with a prograde
orbital geometry is tidally disturbed during the first encounter with
another galaxy and when both galaxies meet again and finally merge
(Mihos \& Hernquist 1996; Murphy et al.\ 2001; Li et al.\
2004).  We choose to model a molecular gas-rich spiral
galaxy on its first encounter with another galaxy of similar mass.

We set up a molecular disk with $M_g=10^{10} \mbox{ M}_{\odot}$ in a
dark matter halo with $M_{halo}=5\times10^{12} \mbox{ M}_{\odot}$.  CO
observations of ULIRGs at both first and second passages show the
presence of molecular disks with $M_g=0.4-1.5\times10^{10} \mbox{
  M}_{\odot}$ (Sanders et al.\ 1988; Solomon et al.\ 1997), which is
in the range found for gas-rich spiral galaxies. However, the
emission originates in regions a few hundred parsecs in radius,
yielding surface densities of $\sim0.5$--$1\times10^4~\mbox{
  M}_{\odot}\mbox{ pc}^{-2}$, within which the molecular mass
dominates the dynamical mass (Sanders et al.\ 1988; Solomon et al.\
1997).  At these high surface densities, molecular hydrogen will
dominate (Blitz \& Rosolowsky 2006), as it can form within a few
million years in turbulent regions with densities over 100~cm$^{-3}$
(Glover \& Mac Low 2007).  The density of H$_2$ traced by CO emission
is $\sim500\mbox{ cm}^{-3}$, comparable to the envelope of giant
molecular clouds, while
a region of much higher density in ULIRGs is traced by HCN emission,
$\sim10^5\mbox{ cm}^{-3}$, comparable to star-forming cloud cores
(Solomon et al.\ 1992).

We assume that the entire interstellar medium (ISM) is a scaled-up version of
a normal galactic disk with the ambient densities 
a factor of $\sim100$ higher, making even the intercloud medium a molecular region. 
Thus we assume that the surface density distribution of the molecular disk is 
exponential, with $\Sigma(R)=\Sigma_0 exp(-R/R_d)$ where $R_d$ is a scale radius
 (see also $\Sigma(R)$ of Arp 220 by Scoville 1997).  

We choose to model a disk with a central surface density of
$\Sigma_{0}=10^4~\mbox{ M}_{\odot}\mbox{ pc}^{-2}$, with a disk scale
radius $R_d=0.7$ kpc. The disk is in hydrostatic equilibrium with a
Navarro, Frenk, \& White (1997; hereafter NFW) halo potential, and a
disk potential based on the thin disk approximation (Toomre 1963),
since $M_{dyn}\approx M_{g}$.  The NFW potential is
\begin{equation}
\label{nfw}
\Phi(x) = \frac{G M_{halo}}{R_v} ~\frac{\ln (1+cx)/x}{F(c)}, 
\end{equation}
where we set the virial radius $R_v = 326$~kpc, $x=r/R_v$, $c$ is a halo
concentration factor, set to $c=5$, appropriate for a large halo
(Jimenez et al.\ 2003), and $F(c)=\ln (1+c)-c/(1+c)$. 

The velocity dispersion of the molecular gas is observed to be
90~km~s$^{-1}$ in Arp 220, which appears to be at the end of the
merging process (Scoville et al.\ 1997).  Such a high velocity dispersion
yields a scale height of 15 pc in its disk (Scoville et al.\ 1997).
Molecular clouds with such high dispersion will get destroyed by
colliding with other clouds, so the cooling time behind the shocks
must be shorter than the destruction time interval.  We assume the gas
is supported by turbulence with a similarly high velocity dispersion
$c_s=55\mbox{ km s}^{-1}$. It is lower than $90\mbox{ km s}^{-1}$
because gravity from our disk gas and halo can not confine the gas
with a higher velocity dispersion.  As a comparison, we also model a
disk with a higher surface density of $\Sigma_0=5\times10^4~\mbox{
M}_{\odot} \mbox{ pc}^{-2}$ with $R_d=0.17$ kpc and $c_s=90\mbox{ km
s}^{-1}$, based on the central surface density observed in Arp 220.
This is the highest $\Sigma_0$ of all observed ULIRGs.

\begin{figure}
\includegraphics[width=\textwidth]{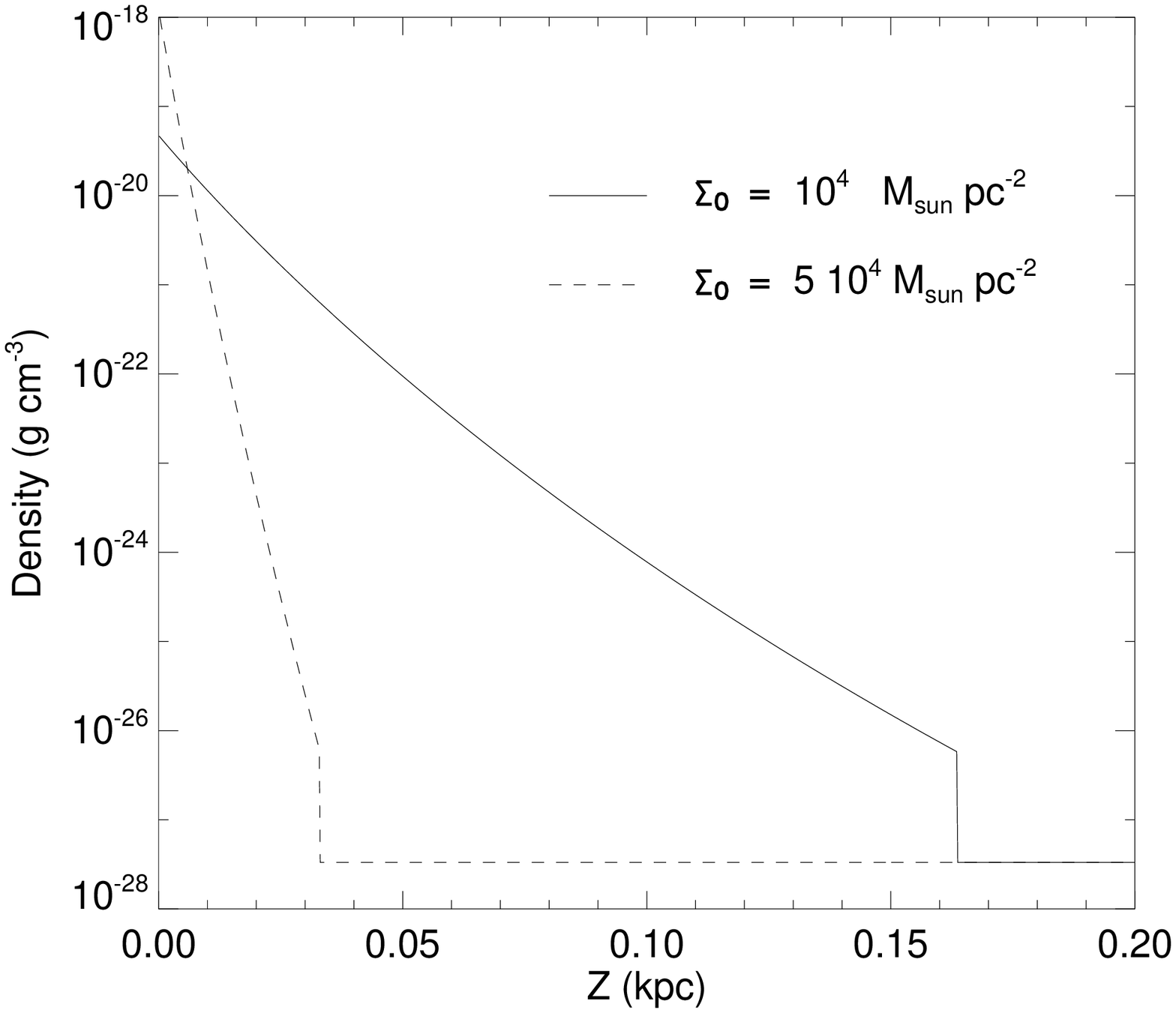}
\figcaption{The density distributions of the disks with central surface 
densities of $\Sigma_0=10^4$ and $5\times10^4~\mbox{ M}_{\odot}\mbox{ pc}^{-1}$
 in the vertical direction. The scale heights are $H=7$ and 2 pc respectively. 
\label{H}
}
\end{figure}

Figure~\ref{H} shows the vertical density distributions of both disks. 
The exponential scale heights are rather small, 7 and 2 pc, but the
gas within them is very  
dense, $\sim5000$ and $10^5\mbox{ cm}^{-3}$ respectively. The gas density is still 
$\sim500\mbox{ cm}^{-3}$ at $Z\ga 4R_d$.
At higher altitudes, where the gas is less dense, the gas is
physically atomic or even ionized. We do not take that state change
into account in our model, though.
When the number density drops to $n=10^{-2}$~cm$^{-3}$ we set the gas
density in the halo constant as it is no longer dynamically important
on the length and time scales treated in our model.

\subsection{Star Formation}
\label{subsec:sf}
We assume a single starburst that occurs at the center of the disk,
and that all the kinetic energy of the starburst supernovae is
released in a central wind of constant mechanical luminosity.  In
reality, the discrete energy inputs from supernovae generate
blastwaves that become subsonic in the hot interior of the bubble
first produced by stellar winds, and hence can be treated as a
continuous mechanical luminosity in the study of bubble dynamics (Mac
Low \& McCray 1988).  These assumptions mean that a single superbubble
forms, evolving to produce a bipolar outflow of gas.

\begin{figure}
\includegraphics[width=0.5\textwidth]{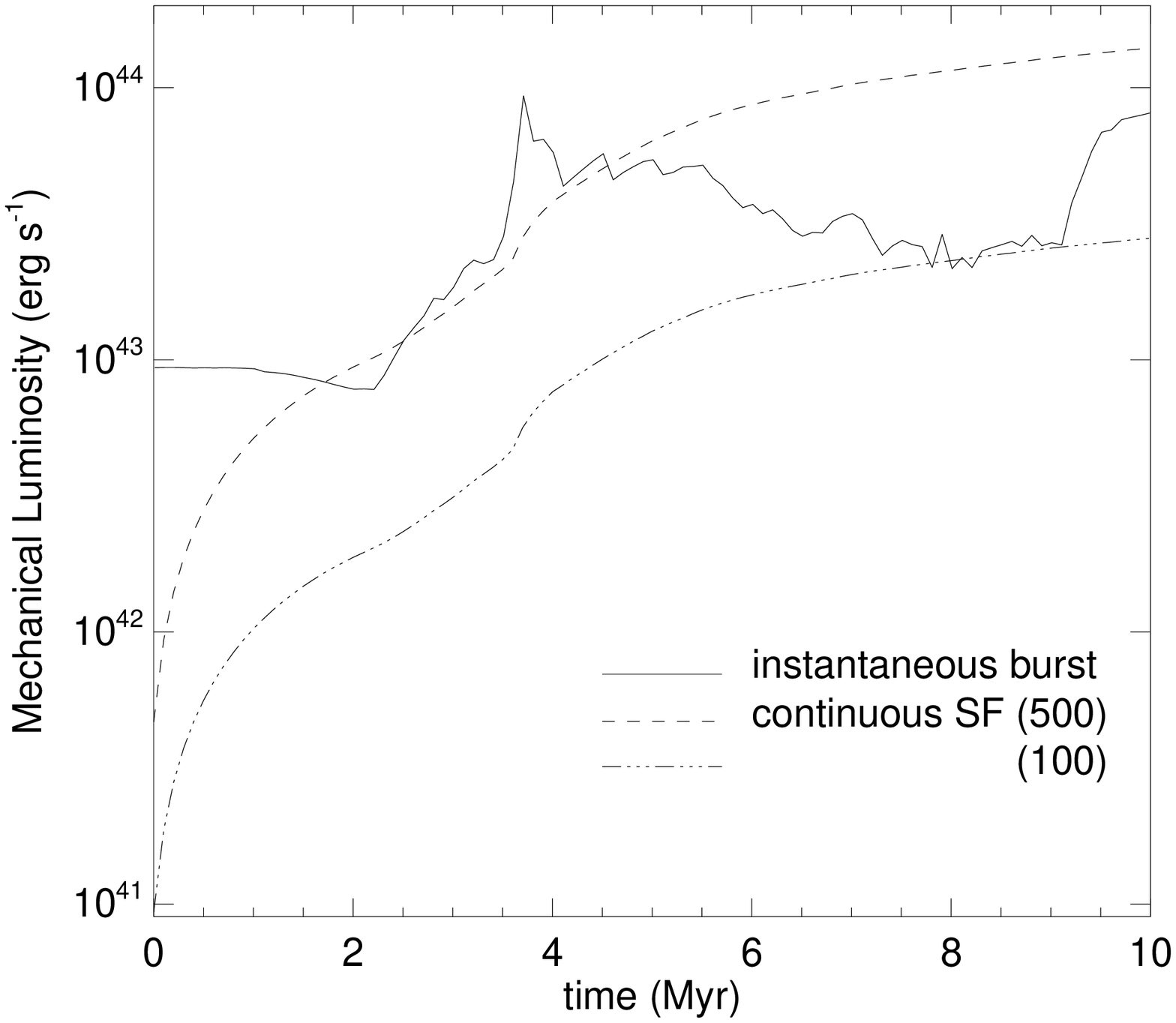}
\includegraphics[width=0.5\textwidth]{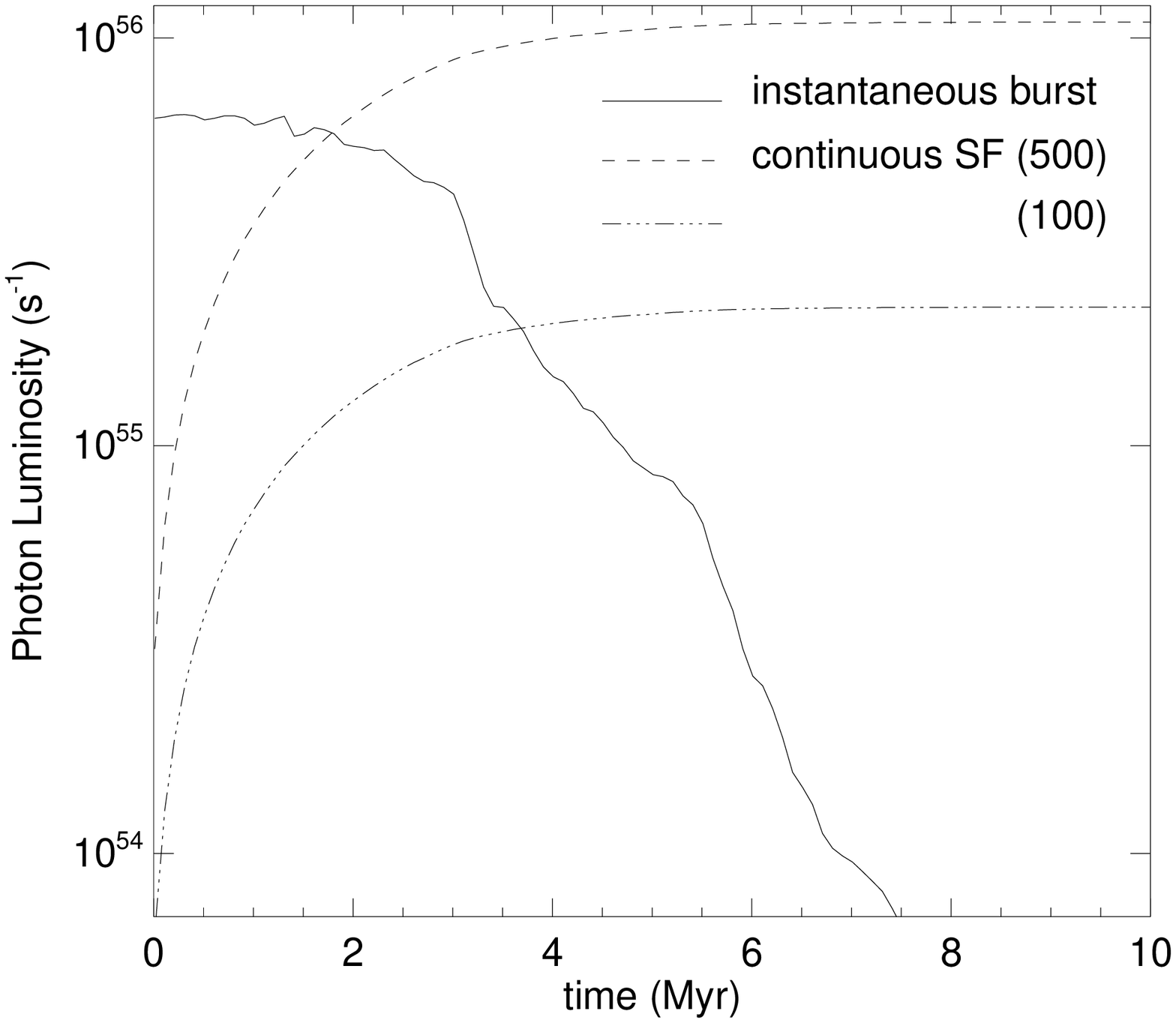}
\figcaption{Mechanical luminosities ({\em left panel}) and ionizing photon luminosity 
({\em right panel}) as a function of time 
for three starburst scenarios: instantaneous starburst with
$M_*=10^9~\mbox{ M}_{\odot}$ ({\em solid line}); continuous star formation
with $500~\mbox{ M}_{\odot}\mbox{ yr}^{-1}$ ({\em dashed line}); and
continuous star formation with $100~\mbox{ M}_{\odot}\mbox{ yr}^{-1}$ ({\em
dash-dotted line}). 
Population synthesis models are from Starburst~99 (Leitherer et al.\ 1999).
\label{lmech}
}
\end{figure}

Figure~\ref{lmech} shows the evolution of mechanical luminosity
$L_{mech}$ as a function of time for an instantaneous starburst with
$10^{9}~\mbox{ M}_{\odot}$ of gas turning into stars and for
continuous starbursts with star formation rates of $100$ and
$500~\mbox{ M}_{\odot}\mbox{ yr}^{-1}$, based on the Starburst 99
model (Leitherer et al.\ 1999).  The Starburst~99 model uses a power
law initial mass function with exponent $\alpha=2.35$ between low-mass
and high-mass cutoff masses of $M_{low}=1~\mbox{ M}_{\odot}$ and
$M_{up}=100~\mbox{ M}_{\odot}$ with solar metallicity.  Star formation
rates in ULIRGs are estimated to be $\ga 100~\mbox{ M}_{\odot}$
yr$^{-1}$ based on far-infrared luminosities and the assumption of
continuous star formation (see Table~1 of Martin 2005; note that the
star formation rates given there correspond to a low mass cutoff of
$0.1~\mbox{ M}_{\odot}$ and must be divided by a factor of 2.55 before
comparison to the population synthesis models).

The amount of mechanical power supplied per unit stellar mass depends
on the star formation history. About $\sim40$ million supernovae, for
example, will be produced by an instantaneous starburst of
$M_*=10^{9}$~M$_{\odot}$.  For a continuous starburst, $L_{mech}$
increases until the death rate of massive stars catches up to their
birth rate, after about 40 Myr. The power rises particularly rapidly
over the first few Myr, the period modeled by our
simulation. Fig.~\ref{lmech}{\em a} shows the evolution for a
continuous starbursts with $500~\mbox{ M}_{\odot}\mbox{ yr}^{-1}$ and
$100~\mbox{ M}_{\odot}\mbox{ yr}^{-1}$.  For our ULIRG models, we use
constant mechanical luminosity winds with values $L_{mech} = 10^{43}$,
$10^{42}$, and $10^{41}\mbox{ erg s}^{-1}$.  The highest of these
corresponds to the mechanical luminosity expected from stellar winds
during the first 2~Myr of an instantaneous starburst with $M_*=10^9
\mbox{ M}_{\odot}$.  The subsequent supernovae will result in a far
higher mechanical luminosity, but that is likely to be vented out of
the galactic disk through the hole opened by the initial
blowout. $L_{mech} = 10^{43}$~erg~s$^{-1}$ also corresponds to a model
with a constant SFR of 15.9 M$_{\odot}$~yr$^{-1}$.  This
correspondence assumes the birth and death rates of massive stars are
in equilibrium, which is achieved about 40 Myr after the burst begins.
In this scenario, the burst would been ongoing through its initial
stages before our simulation starts, but gas displaced by the initial
feedback had been replaced by inflows, and prior feedback energy was
largely radiated.  Our simulation is not such a good representation of
this scenario.

We note it is an oversimplification to assume a single starburst at
the disk center for modeling a galactic outflow. Bipolar outflows are
seen in some starburst galaxies such as M82 (Strickland \& Stevens
2000; Strickland et al.\ 2004) for example, although some ULIRG winds
appear to require starburt regions extended to $\ga1$ kpc to launch
the cool outflow (Martin 2006).  3D, hydrodynamic simulations of dwarf
starbursts have shown that extended, multiple energy sources, as well
as a single central energy source, form a bipolar outflow (Fragile et
al.\ 2004).  These simulations also demonstrated that the main effect
of multiple sources was to reduce the fraction of metals and energy
ejected from the galaxy from almost unity to around 50\%.  Our
assumption thus represents a reasonably strong {\em lower} bound to
the amount of kinetic energy that will be deposited in the observed
cold gas.  We therefore start with this assumption and do not expect
the results to differ much from those expected with more extended star
forming regions.

In addition, we neglect the effects of UV radiation on molecular
hydrogen in the disk.  The UV radiation from massive stars may
photo-dissociate some of molecular hydrogen outside star-forming cores
to atomic hydrogen.  However, the assumed turbulent pressure with
$c_{s}=55\mbox{ km s}^{-1}$ is 14 times greater than the increased
thermal pressure by photo-dissociation.  We thus safely neglect the
effects of UV radiation on the disk gas structure.

We define the model with $\Sigma_0=10^4~\mbox{ M}_{\odot}\mbox{
pc}^{-2}$ and $L_{mech}=10^{43}\mbox{ erg s}^{-1}$ as our fiducial
model (U1/X1). We list the parameters for all the other runs in
Table~\ref{run}.

\subsection{Bubble Dynamics}

In the molecular disk of a ULIRG with a very small scale height, a
bubble with $L_{mech}\ge10^{41}\mbox{ erg s}^{-1}$ quickly blows out
of the disk at $t\ll 1$ Myr.  We can only simulate the evolution of
the bubble up to $t\sim0.3-1$ Myr before it leaves our grid, which is
rather small because of the cost of high resolution. We argue in
\S~\ref{sec:observation} that cooled, swept-up shells, which we
identify as Na~{\sc i} absorbing gas, can acquire maximum velocities
primarily determined by when they blow out, accelerate, and fragment.
The question remains whether these shell fragments and clumps survive
as the hot interior wind streams through them.

Each individual shell fragment after blowout remains unstable to
smaller scale R-T instabilities while being further accelerated by the
wind, requiring extremely high resolution to fully resolve.
Resolution is not as critical for previously published numerical
studies on starburst winds that explored feedback parameters.  For
example, reasonable assumptions about cooling losses indicate that
moderate luminosity starbursts do not remove a significant fraction of
the galaxy's gas from the halos (Mac Low \& Ferrara 1999), and the
mass loss rates would only decrease further with more cooling.  A
large fraction of the heavy elements do escape from the halos (Mac Low
\& Ferrara 1999; Fujita et al.\ 2004), and this result does not depend
on resolution below $\sim$ a few tens of parsecs, so long as at least
a few R-T modes are excited in the swept-up shells to let the
metal-enriched gas escape.  Therefore, a major question that must be
answered to understand observations of cold gas at high velocity is
whether clumps of dense gas survive in the lower density wind (cf.\
Klein et al.\ 1994; Marcolini et al.\ 2005), 
  where they will fragment because of
hydrodynamic instabilities, such as R-T (strictly speaking,
Richtmyer-Meshkov; see Richtmyer 1960; Meshkov 1969), and Kelvin-Helmholtz.

\begin{figure}
\includegraphics[width=0.5\textwidth]{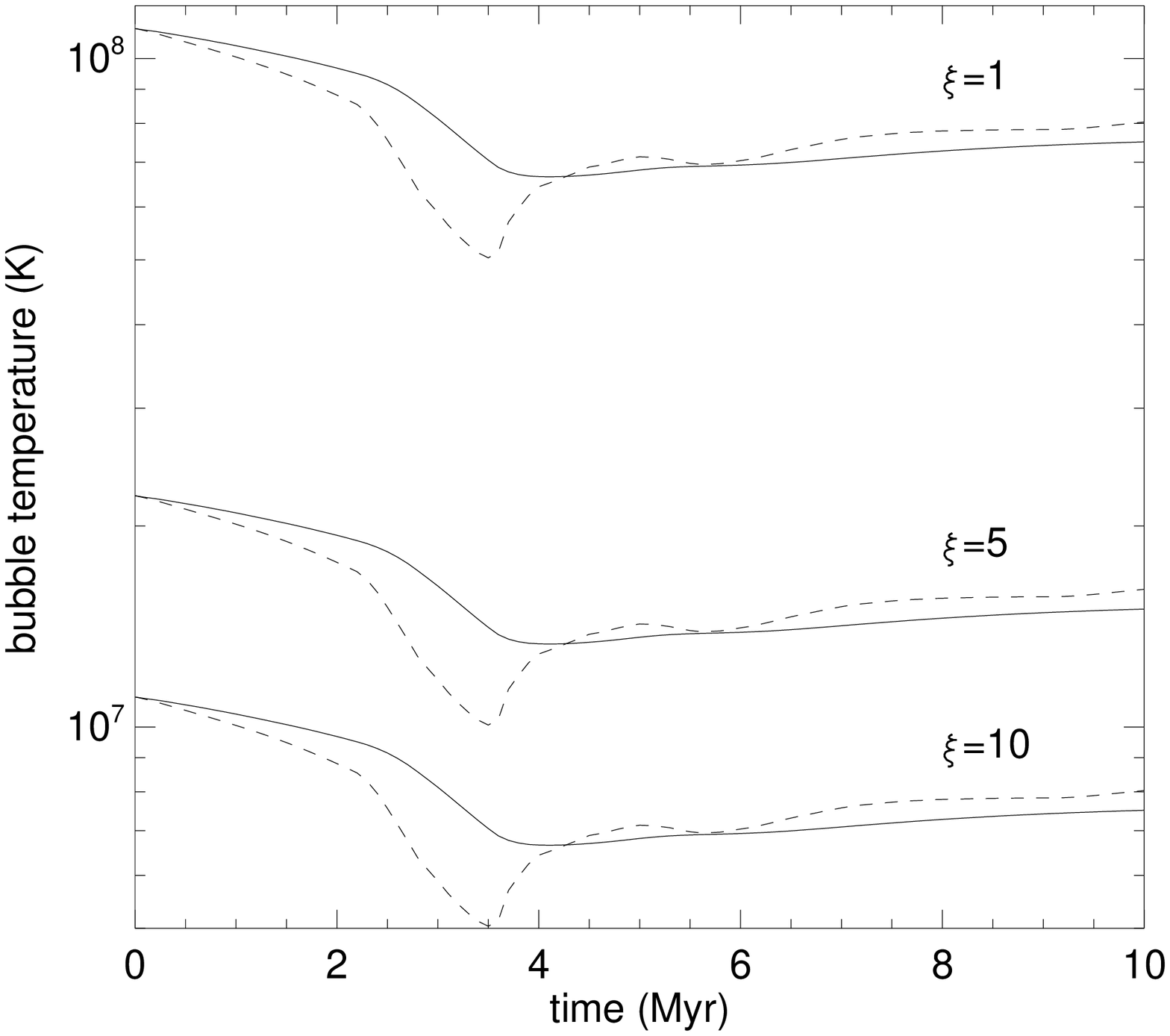}
\includegraphics[width=0.5\textwidth]{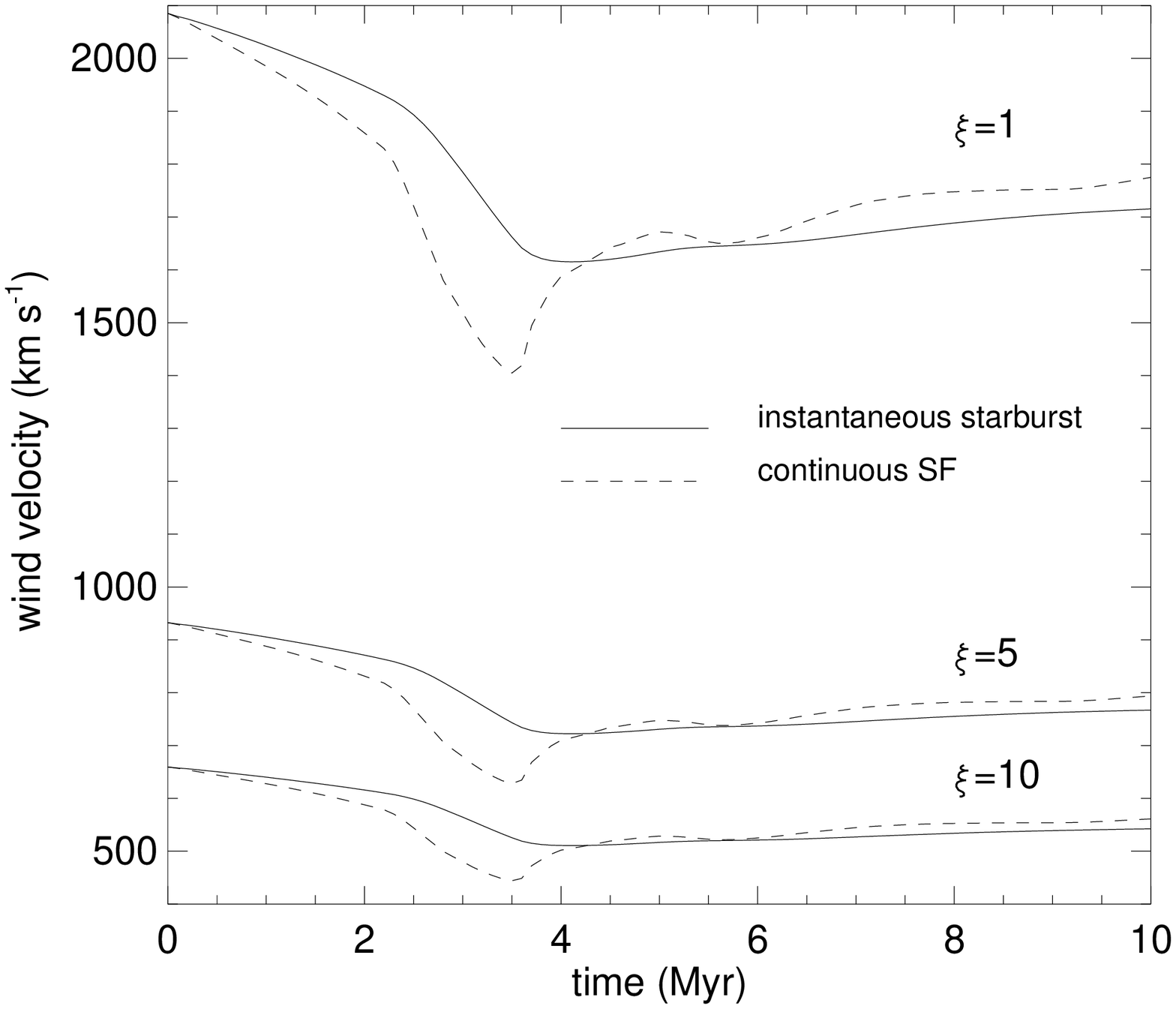}
\figcaption{The ({\em a}) temperature of a hot bubble in
  a uniform medium driven by a starburst 
  with 
$L_{mech}$ and the rate of supernova ejecta $M_{SN}\propto L_{mech}$ predicted by the Starburst 
99 model as a function
  of time, with the given mass-loading factors $\xi$, under the
  assumption of an instantaneous ({\em solid line}) or continuous
  ({\em dashed line}) starburst.  The amount of mass in the hot wind
  is $\xi M_{SN}$. ({\em b}) The expected terminal velocity of the
  wind driven by such a bubble after its blowout from a stratified
  disk. Note $T_{wind}$ and $v_{wind}$ are proportional to $L_{mech}/M_{SN}$, thus 
  independent of SFR assumed. 
\label{vtwind}
}
\end{figure}

Weaver et al.\ (1977) argued that the density in the hot interior of
bubbles in uniform gas is dominated by conductive evaporation from the
dense shell.  However, our model does not explicitly include thermal
conduction, or material ablated off of high-density molecular clouds
associated with the central starburst.  Instead, we add additional
mass to our central luminosity source to account for this process,
multiplying the mass input rate by a mass-loading factor $\xi$.  The
amount of thermally evaporated mass in a bubble expanding into a
uniform medium is proportional to $L^{27/35}_{mech}\rho^{-2/35}$.  
The internal temperature is
\begin{equation}
T_b(t)=(\gamma-1)\frac{\mu}{k_{B}}
\left[\frac{5}{11}\int_{0}^{t}L_{mech}(t') dt' \right] / 
\left[\int_0^t\xi M_{SN}(t') dt' \right].
\end{equation}
where $L_{mech}(t)$ and the mass of supernova ejecta $M_{SN}(t)$ as a
function of time are taken from Starburst 99 model, the adiabatic
index $\gamma=5/3$, mean mass per particle $\mu=14/22 m_H$, and $k_B$
is the Boltzmann factor.  Weaver et al.\ (1977) showed that $5/11$ of
the total input mechanical energy goes into the hot, pressurized
region bound by the inner and outer shock fronts (see their eq.~[14]).
This was applied to superbubbles by McCray \& Kafatos (1987).  We show
$T_b(t)$ in FigMac Low \& McCrayure~\ref{vtwind}{\em a}.

We can estimate the terminal velocity of the wind driven by such a
bubble by equating the bubble interior energy with the kinetic energy
of the mass-loaded wind, so that
\begin{equation}
\label{ewind}
v^2_{wind}(t)=2\left[\frac{5}{11}\int_0^t L_{mech}(t')
  dt'\right]/\left[\int_0^t \xi  M_{SN}(t')dt'\right],
\end{equation}
which we plot in Figure~\ref{vtwind}{\em b}.  Since the mechanical
luminosity $L_{mech}(t)$ and the mass of supernova ejecta $M_{SN}(t)$
are both linearly proportional to the amount of gas converted to
stars, $T_{b}$ and $v_{wind}$ are the same for all strengths of
starburst with the same mass-loading rate $\xi$.  We take a fiducial
$\xi$ value of about 8 corresponding to a mass-loading rate
$\dot{M_{in}}=17 (L_{mech}/10^{43})~\mbox{ M}_{\odot}\mbox{ yr}^{-1}$,
but run test simulations with $\xi$ varying between $\sim$1--15 to
explore the effects of mass-loading on the shell
kinematics. Comparisons to observations suggest $\xi\approx10$
(Suchkov et al.\ 1996 -- see M82 comparison; Martin et al.\ 2002).
Throughout this paper, we designate as the wind the hot interior gas
freely streaming outwards after blowout of the shell.

\section{Numerical Methods}
\label{sec:numerics}

We follow the numerical methods used by Mac Low \& Ferrara (1999) and
Fujita et al.\ (2003) to model a starburst in a galactic disk.  We
briefly summarize our methods below, but refer to the papers above for
more details. We compute the evolution of a starburst-driven blastwave
as it blows out of the molecular disk of a ULIRG with
ZEUS-3D\footnote{Available from the Laboratory for Computational
Astrophysics, http://cosmos.ucsd.edu}, an Eulerian, finite-difference,
astrophysical gas dynamics code (Stone \& Norman 1992a; Clarke 1994),
that uses second-order van Leer (1977) advection, and a quadratic
artificial viscosity to resolve shock fronts. We use the loop-level
parallelized version ZEUS-3D, in its 2D form.  Runs were
done on Silicon Graphics Origin 2000 machines using eight processors,
and typically took $\sim$1--12 days.

We assume azimuthal symmetry around the rotational axis of the galaxy.
Our fiducial grids are $1000\times500$ zones with a resolution of 0.2
pc, comparable to the size of star-forming cloud cores. We also run
the same simulations with resolution of 0.1, 0.4, and 0.8 pc to
examine the sensitivity of post-shock shell density and thus R-T
instability growth to resolution. We use reflecting boundary
conditions along the symmetry axis and along the galaxy midplane and
outfall boundary conditions on the other two axes.

The assumption of azimuthal symmetry limits Rayleigh-Taylor
instabilities to growing as rings, reducing the number of clumps below
what would actually be expected from 3D spikes. Mac Low et al.\ (1989)
compared models with azimuthal to slab symmetry to demonstrate that
the fixing of a central axis of symmetry did not markedly change the
behavior. 3D models of isolated superbubbles have only been performed
for the magnetized case starting with the work by Tomisaka
(1998). Recently Stil et al.\ (2008) have performed 3D studies of
unmagnetized superbubbles as calibration for a study of magnetized
superbubbles, but they did not extend their hydrodynamical models into
the R-T unstable regime. Comparison of 2D to 3D models of shocked
clouds by Stone \& Norman (1992b) and Xu \& Stone (1994) found little
difference in their dynamical evolution aside from the breakup of
post-shock vortex rings in 3D. Young et al.\ (2001) and Cabot (2006)
compared high resolution 2D and 3D models of planar, incompressible
R-T instability.  Cabot (2006) cautions that 2D models produce larger,
less-well mixed structures at late times because of the inverse energy
cascade that occurs in 2D flows.

To drive a constant luminosity wind, we add mass and energy to a
source region with a radius of 10 pc (50 zones).  Our fiducial mass
input rate is $\dot{M_{in}}=17(L_{mech}/10^{43})~\mbox{
M}_{\odot}\mbox{ yr}^{-1}$ which corresponds to a mass-loading factor
$\xi\approx8$.  For simulations with different resolutions, we keep
the the number of zones covering the spherical edge of the source
regions the same by maintaining its radius as a constant number of
zones rather than a constant physical size.  This is important because
aliasing at the edge creates density perturbations that can be
amplified by hydrodynamic instabilities, such as the R-T instability
in the swept-up shell. We use ratioed grids for lower-resolution runs
and we decrease the size of source region to a radius of 5 pc (50
zones) for a higher-resolution run.  We directly show the effects of
this initial noise on the development of R-T instability by running a
simulation with a resolution of 0.2 pc, but with a source region with
a radius of 5 pc (25 zones).

As in Mac Low \& Ferrara (1999), we use a cooling curve by MacDonald \& Bailey (1981) for
solar metallicity with a temperature floor of either $T_{floor}=10^2$
K, the temperature to which the metals can cool the gas, or $10^4$ K,
the temperature maintained by photoionization heating.  The shocked
gas in swept-up shells cools efficiently to the temperature floor set
in the cooling curve, since our molecular disk is very dense.  We show
in \S~\ref{subsec:effects-cc} that even our highest resolution runs do
not yet fully resolve the dense shells even for $T_{floor} = 10^4$~K,
so the influence of the cooling floor is not evident in our work.  We
also include an empirical heating function tuned to balance the
cooling in the background atmosphere. This is linearly proportional to
density, so that it is overwhelmed by cooling in compressed gas which
is proportional to the square of the density (Mac Low et al.\ 1989).
This is to prevent the background atmosphere from spontaneously
cooling.  We use a tracer field (Yabe \& Xiao 1993) to turn off
radiative cooling in the hot bubble interior, in order to prevent mass
numerically diffused off the dense shell from spuriously cooling the
interior. The cooling time of the interior is much longer than the
dynamical time of our bubble, so interior cooling is physically
unimportant to the bubble dynamics (Mac Low \& McCray 1988).
These adiabatic bubble interiors form energy-driven winds after blowout.

\section{Parameter Studies}

We now describe the results of parameter studies of both physical and
numerical variables. We begin by considering physical variables,
including mechanical luminosity, mass-loading of the wind, and disk
surface density.  We then discuss numerical variables, focusing on
how numerical resolution and the cooling cutoff temperature affect
shell density and also examining the effect of changing the size of
the source region.

\subsection{Physical Parameters}
\subsubsection{Mechanical Luminosities}
Our fiducial model (X1) has mechanical luminosity
$L_{mech}=10^{43}\mbox{ erg s}^{-1}$.  This model corresponds to the
first 2~Myr of a starburst in which $10^9~\mbox{ M}_{\odot}$ of gas
turns into stars instantaneously. We compare this to models with lower
mechanical luminosities $L_{mech}=10^{42}$ and $10^{41}$ erg s$^{-1}$
(models X2 and X3 respectively) in the same molecular disk.  These
mechanical luminosities correspond to instantaneous bursts of $10^8$
and $10^7$~M$_{\odot}$.  Figure~\ref{L43} shows the density
distribution of our fiducial model in its right panel.  This may be
compared to Figure~\ref{L421}, which shows the density distributions
of the two models with lower $L_{mech}$ at $t\approx0.49$ Myr and 0.85
Myr respectively.  The swept-up shells fragment due to R-T instability
into multiple clumps and shells.

Secondary Kelvin-Helmholtz instabilities ablate the sides of these
fragments as the hot gas streams through them. Look at the clumps, for
instance, at (R,Z)$\approx(40,65)$, (30,110), and (15--25,120) pc in
X1.  Note also that the swept-up shells in the horizontal direction
are also R-T unstable, because our disk gas is stratified in the
radial direction, too, due to its exponential surface density profile.

In fact, the degree of fragmentation is larger in X2 and more so in X3
because $\dot{M_{in}}\propto L_{mech}$ and so the density of interior
gas is lower.  In particular, most shell fragments in X3 are already
falling back to the disk.  A mechanical power of
$10^{42}$~erg~s$^{-1}$ is just too small in such a dense environment
to accelerate the bulk of the shells to the disk's escape velocity.

\subsubsection{Surface Density}
The top right panel of Figure~\ref{L421} shows the density
distribution of our model with $M_d=10^{10}~\mbox{ M}_{\odot}$, but
with a higher surface density, $\Sigma_0=5\times10^4~\mbox{
M}_{\odot}\mbox{ pc}^{-2}$ at $t=0.22$ Myr (V1).  With the same
mechanical luminosity, $L_{mech}=10^{43}\mbox{ erg s}^{-1}$, the
bubble blows out earlier at $t\approx0.15$ Myr, because the disk is
more stratified with a smaller scale height (see Figure~\ref{H}).
Except the time of blowout, the degree of fragmentation and the shell
kinematics are about the same in models with surface
densities different by a factor of five.

\subsubsection{Mass-Loading}

Mass-loading from thermal conduction and molecular clouds determines
the density of the bubble interior and wind.  Figure~\ref{w} shows the
density distributions of our fiducial ULIRG model with
$T_{floor}=10^2$ K with mass-loading rates of 1.7, 17, 49, and 120
$\mbox{ M}_{\odot}\mbox{ yr}^{-1}$ (models U1-A, U1, U1-B, and U1-C).
These mass-loading rates correspond to bubble interior temperatures,
$T_{b}=1.7\times10^8$, $2.2\times10^7$, $7.5\times10^6$, and
$3.2\times10^6$ K and wind terminal velocities expected when all the
thermal energy is converted to kinetic energy, $v_{wind}=2700$, 1000,
600, and 250 km s$^{-1}$ respectively.

Figure~\ref{w} shows the bubbles just before they leave the grid, at
$t=0.23$, 0.27, 0.35, and 0.41 Myr respectively.  Since the input
mechanical luminosity is the same in all these models the bubbles
initially grow at about the same rate, driven by the thermal pressure
of the hot interior gas.  However, a bubble with a lower mass-loading
rate and higher wind terminal velocity expands faster into the halo
once the swept-up shells fragment and the hot gas blows out between
the fragments.  In addition, the blowout occurs earlier and in more
places with a lower mass-loading rate because the density of the
bubble interior gas is lower.  A higher density contrast between the
hot interior and the swept-up shells promotes shell fragmentation by
R-T instability, the growth of which is proportional to the density
contrast.  In the least dense model U1-A, all the swept-up shells
quickly fragment into fingers and filaments and the dense clumps at
their edges are subject to secondary Kelvin-Helmholtz instabilities.

As the hot bubble interior becomes transonic after blowout, it
accelerates the shells and shell fragments by ram pressure rather than
thermal pressure.  This low-density wind thus can accelerate the
shells to higher velocity after blowout. We quantitatively study the
effects of wind ram pressure in the next section
(\S~\ref{subsec:AMvelocity}).  The observed X-ray temperature $T_X$ is
$\sim0.67$ keV $ = 7.7 \times 10^6$~K in all kinds of starburst
galaxies from dwarfs to ULIRGs (Martin 1999; Heckman et al.\ 2001; Huo
et al.\ 2004; Grimes et al.\ 2005).  This corresponds to
$\xi\approx10$ in equation~\ref{ewind}.  However, the X-ray emission
is proportional to $n^2$ so it is biased toward high-density regions
such as the interface between the hot interior gas and the shells and
their fragments.  At this interface, conductive evaporation and
turbulent ablation raise the density.  The recent observations of
diffuse hard X-ray emission in starburst galaxies suggest the
existence of a very hot ($\log T\ga7.5$) metal-bearing gas
(e.g. Strickland et al.\ 2004; Strickland \& Heckman 2007).  
Then the bulk of the hot wind may
still be very hot $\sim10^8$ K, the temperature which Strickland \&
Stevens (2000) modeled for M82 with $\xi=1$.

\subsection{Numerical Parameters}

\subsubsection{Grid Resolution}
\label{subsec:effects-res}

Figure~\ref{res} shows the density distributions of our fiducial model
with grid resolution varying from 0.1 to 0.8~pc.  We chose
$T_{floor}=10^4$ K for this resolution study, because shell densities
are not too far from what we expect analytically with this high
minimum temperature.  The growth of R-T instability is significantly
enhanced in the highest resolution run, X1-0, and suppressed in the
lower resolution runs, X1-2 and X1-4. In particular, all the outermost
shells seen in X1 are further fragmented by R-T instability in X1-0
with the resolution increased by a factor of only two.  The positions
of outer shock fronts in the horizontal direction agree very well
among the four simulations, since the shells there are not subject to
severe hydrodynamic instabilities.

Figure~\ref{dres} shows the density profiles at the outer shock fronts
before any fragmentation occurs for the runs in our standard
resolution study.  The bubble in our highest resolution run X1-0 grows
a little slower in the beginning, because we chose the size of the
source region in X1-0 to be half of that in X1 in order not to
overproduce noise at the contact discontinuity.  We will show below
that this noise feeds R-T instability and must be maintained the same
in order to study the effects of resolution on shell fragmentation
alone.  Thus the density profile of X1-0 at $t=0.06$ Myr is not
directly comparable to those of other runs at $t=0.05$ Myr, but
Figure~\ref{dres} still demonstrates the trend in resolving the peak
shell density as a function of resolution.

Figure~\ref{dres} shows that the shell density is progressively better
resolved as the resolution increases.  However, the cooling times at
the shock front are typically of order $10^2$~yr (see
\S~\ref{subsec:effects-cc}), so a shock with cooling floor equal to
the background temperature of $10^4$~K may be treated as
isothermal. At the time displayed, the Mach number of the outer shock
is ${\cal M}=19$. The shell density we expect from an isothermal shock
propagating into a background density $\rho_{bg}$ is $\rho_{bg} {\cal
M}^2 = 6.1\times10^{-19}\mbox{ g cm}^{-3}$.  Figure~\ref{dres} shows
that the shell density $\rho_{shell}$ is still unresolved by a factor
of $\sim4$ even with our highest grid resolution of 0.1 pc.

The difference we see in the development of R-T instability develops
because of two factors.  First, the post-shock density of the swept-up
shells is not fully resolved. Increasing density contrast drives
faster R-T growth.  Second, the linear R-T instability grows more
quickly at smaller wavelengths, but the nonlinear development moves to
increasing larger wavelengths as competition between growing modes
becomes important (Youngs 1984).  At least 10--25 zones is required to
resolve the smallest modes, though (e.g. Mac Low \& Zahnle 1994), so
grid resolution matters critically for the initial development and
transition to nonlinearity.  However, we believe that we reached the
point where resolution effects are no more important than physics we
have not included such as magnetic fields, non-equilibrium cooling,
thermal conduction, and photoionization, as well as the assumption of
azimuthal symmetry (see \S~\ref{sec:numerics}).

For example, strong magnetic fields $B\sim20\mu G$ observed in the
Antennae merging galaxy (Chy\.zy \& Beck 2004) can potentially inhibit
the formation of cold, dense shells or suppress their
fragmentation. Our study does show that the degree of R-T
fragmentation is important to reproduce the observed wide range of
cooled shell fragments. Recall, however, that the shell density in our simulations
can be more than an order of magnitude underresolved. Thus the degree
of fragmentation will not be significantly overestimated in our
simulations unless the magnetic fields are strong enough to reduce the
shell density by more than that.

The density $\rho_{sh}$ expected behind an isothermal shock running
into a magnetic field with Alfvenic Mach number $\cal{M_A}$ and having
$1 \ll {\cal M}_A \ll {\cal M}$ is $\rho_{sh} = \rho_0
\sqrt{2}{\cal M}_A$ (Draine \& McKee 1993). Before the effects of
magnetization become important in limiting R-T instability, the ratio
between the magnetized and unmagnetized post-shock densities must be
of order $2^{1/2} {\cal M}_A / {\cal M}^2 < 10^{-2}$.  With
$\rho\sim10^{-24}\mbox{ g cm}^{-1}$ and $v_s=500-800\mbox{ km
s}^{-1}$, the shell density expected with $B \sim 20 \mu$G becomes
three order of magnitude lower than that without B.  The effect of
magnetic field thus becomes substantial only when a bubble grows to
the high-Z, low-density part of the disk, above $Z>100$ pc.  However,
most of the fragmentation occurs within $Z\sim100$ pc especially with
the highest-resolution run, Thus our results on the wide absorption
profiles are robust.  If anything, the suppression of fragmentation by
magnetic pressure at high latitude will allow the hot wind to keep
accelerating the outermost shells.  Magnetic pressure might thus even
increase the amount of cool gas with high terminal velocity.
 
We also note that resolving the shell density profile is not important
to following the overall dynamical evolution of bubbles driven by the
thermal energy of the hot, pressurized regions (Castor et al.\ 1975;
Weaver et al.\ 1977), but is important to follow the details of shell
fragmentation due to R-T instability and the fate of dense fragments
and clumps by shocks and following hydrodynamic instabilities.  We
will show below that the resolution of 0.2 pc is still not sufficient
to properly model hydrodynamic instabilities acting on shells and
clumps, but is just sufficient for the purpose of demonstrating a wide
range of velocities in shell fragments caused by R-T instability.

\subsubsection{Cooling Temperature Floor}
\label{subsec:effects-cc}
\begin{figure}
\includegraphics[width=\textwidth]{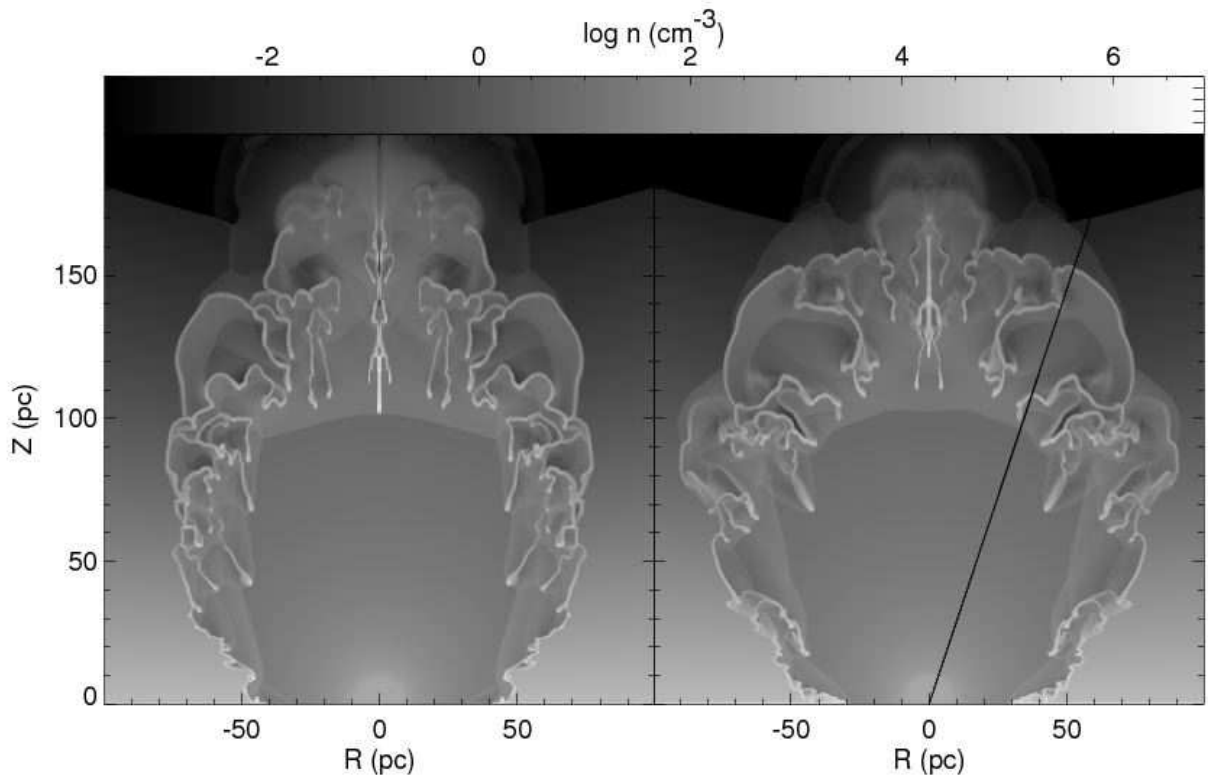}
\figcaption{The density distributions of our standard model at
$t=0.27$ Myr with the cooling temperature cut-offs of $T_{floor}=10^2$
(U1: {\em left panel}) and $10^4$ K (X1: {\em right panel})
respectively. Note that because the shell remains underresolved, the
temperature floor does not substantially influence the behavior of the
shell in our models.  The dark line denotes our fiducial line of
sight.
\label{L43}
}
\end{figure}

\begin{figure}
\includegraphics[width=\textwidth]{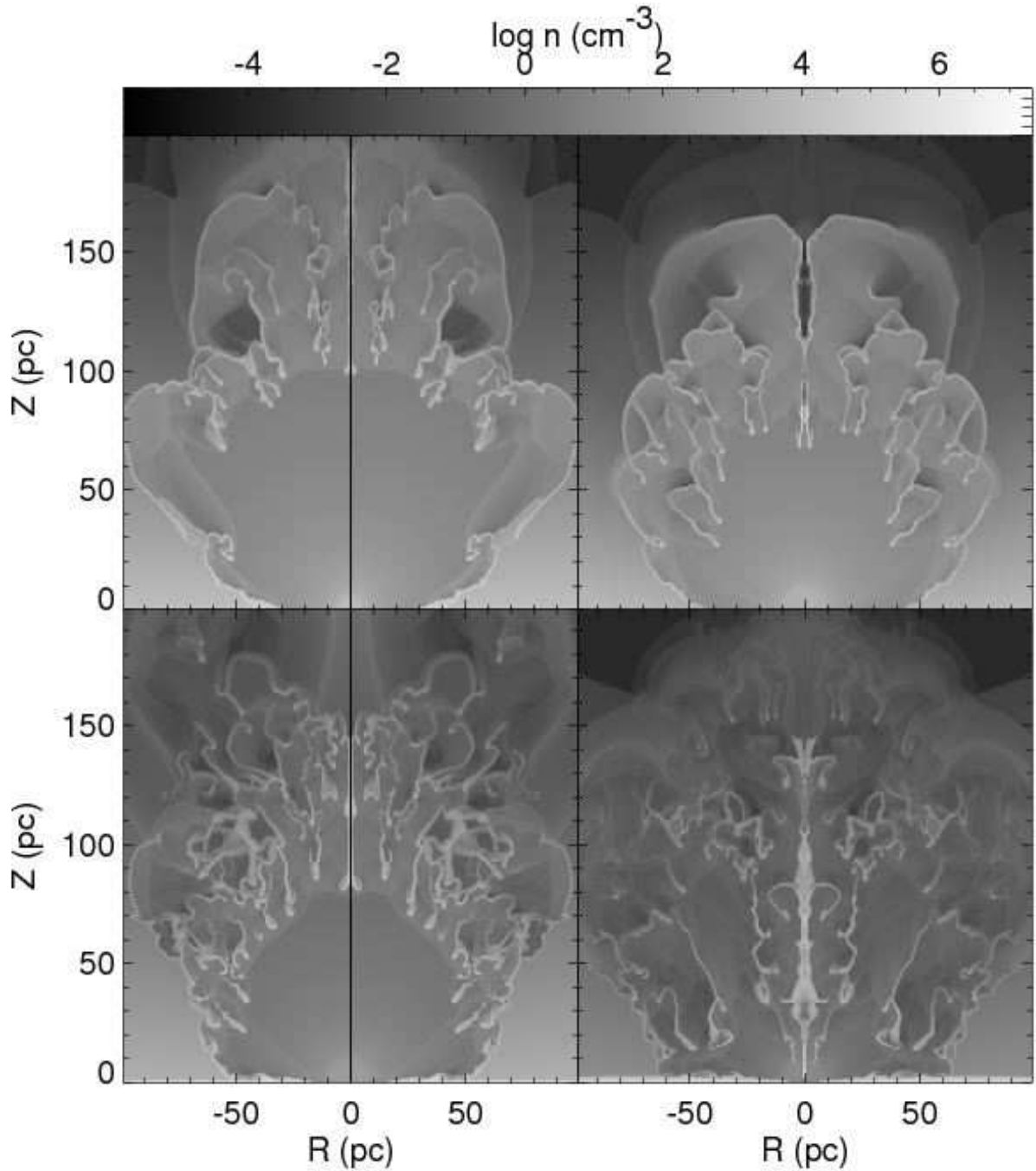}
\figcaption{
({\em top}) The density distributions of two models with
$L_{mech}=10^{43}\mbox{ erg s}^{-1}$: ({\em top left}) our fiducial model with a
source region of only 25 zones, model S1, and ({\em top right}) model V1 with a
higher surface density $\Sigma_0=5\times10^4~\mbox{ M}_{\odot}\mbox{
  pc}^{-2}$. ({\em bottom})The density distributions of two models with lower mechanical
luminosities:  ({\em bottom left}) model X2 with 
$L_{mech}=10^{42}$ and ({\em bottom right}) model X3 with
$L_{mech}=10^{41}\mbox{ erg s}^{-1}$.
Each model is shown just before it exits the grid, at times of $t=0.28$, 0.22, 0.49, and 
0.85 Myr respectively.  
\label{L421}
}
\end{figure}

\begin{figure}
\includegraphics[width=\textwidth]{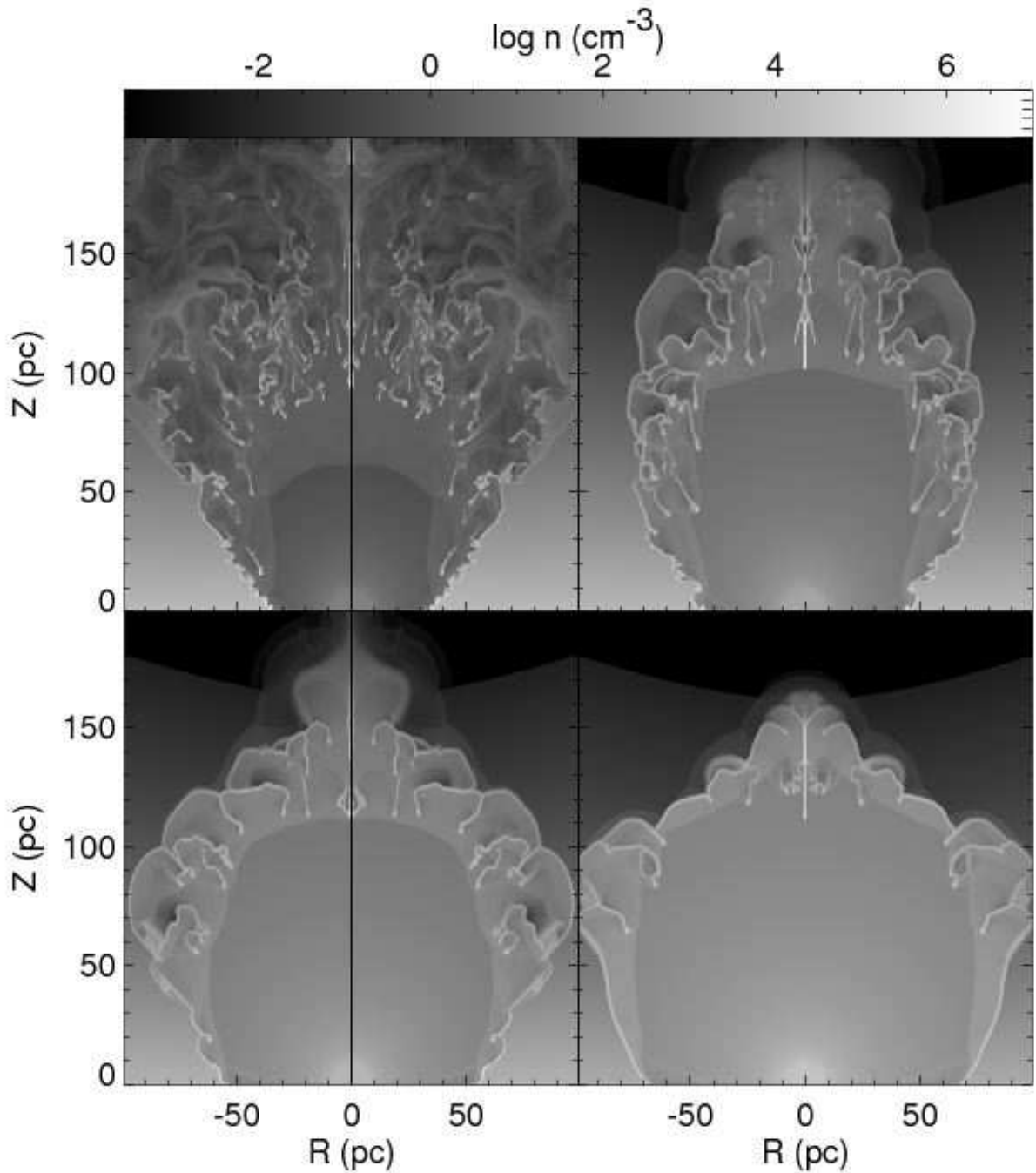}
\figcaption{The density distributions of our model with $T_{floor}=10^2$ K 
with different mass-loading rates: $\dot{M_{in}}=1.7 
\mbox{ M}_{\odot}\mbox{ yr}^{-1}$ (U1-A: {\em top left}),  
17 $\mbox{ M}_{\odot}\mbox{ yr}^{-1}$ (U1: {\em top right}), 49 $\mbox{
  M}_{\odot}\mbox{ yr}^{-1}$(U1-B: {\em bottom left}), 
and 120 $\mbox{ M}_{\odot}\mbox{ yr}^{-1}$ (U1-C: {\em bottom right}). They are shown at $t=0.22$, 0.27, 0.35, 
and 0.41 Myr respectively. 
Different mass-loading rates correspond to different wind temperatures and terminal wind velocities. 
 Note that the growth of hydrodynamic instabilities is suppressed 
as the mass-loading rate decreases and the terminal wind velocity increases. 
\label{w}
}
\end{figure}

\begin{figure}
\includegraphics[width=\textwidth]{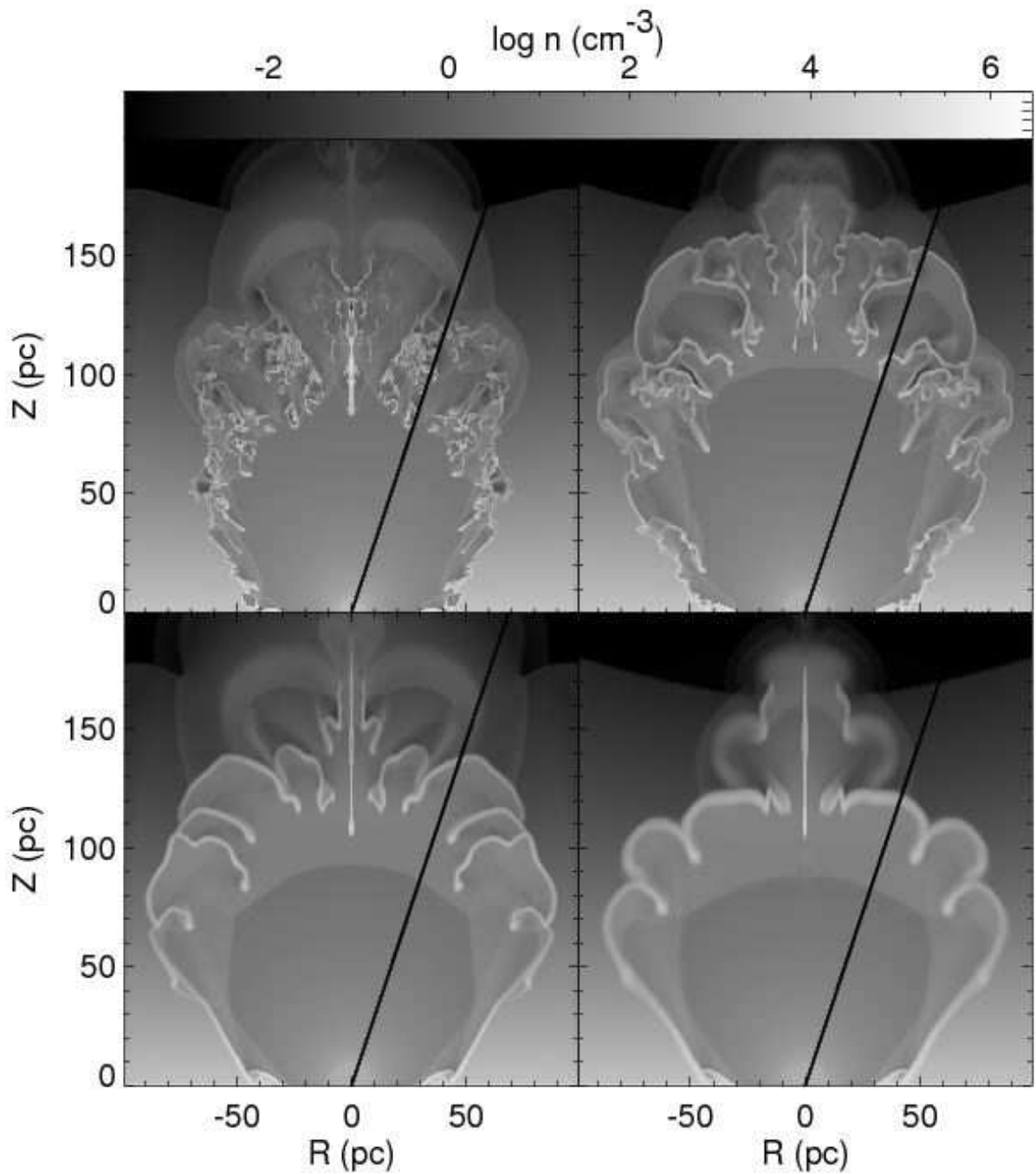}
\figcaption{The density distributions of our standard model with $T_{floor}
=10^4$ K at $t=0.27$ Myr, with 
resolution of 0.1 pc (X1-0: {\em top left}), 
our fiducial
resolution of 0.2 pc 
(X1: {\em top right}), and 
resolutions of 0.4 pc (X1-2: {\em bottom left}) 
and 0.8 pc (X1-4: {\em bottom right}). Note that the growth of 
R-T instability is suppressed as the resolution decreases. The black
lines show the typical line of sight used for line profile analysis.
\label{res}
}
\end{figure}

\begin{figure}
\includegraphics[width=\textwidth]{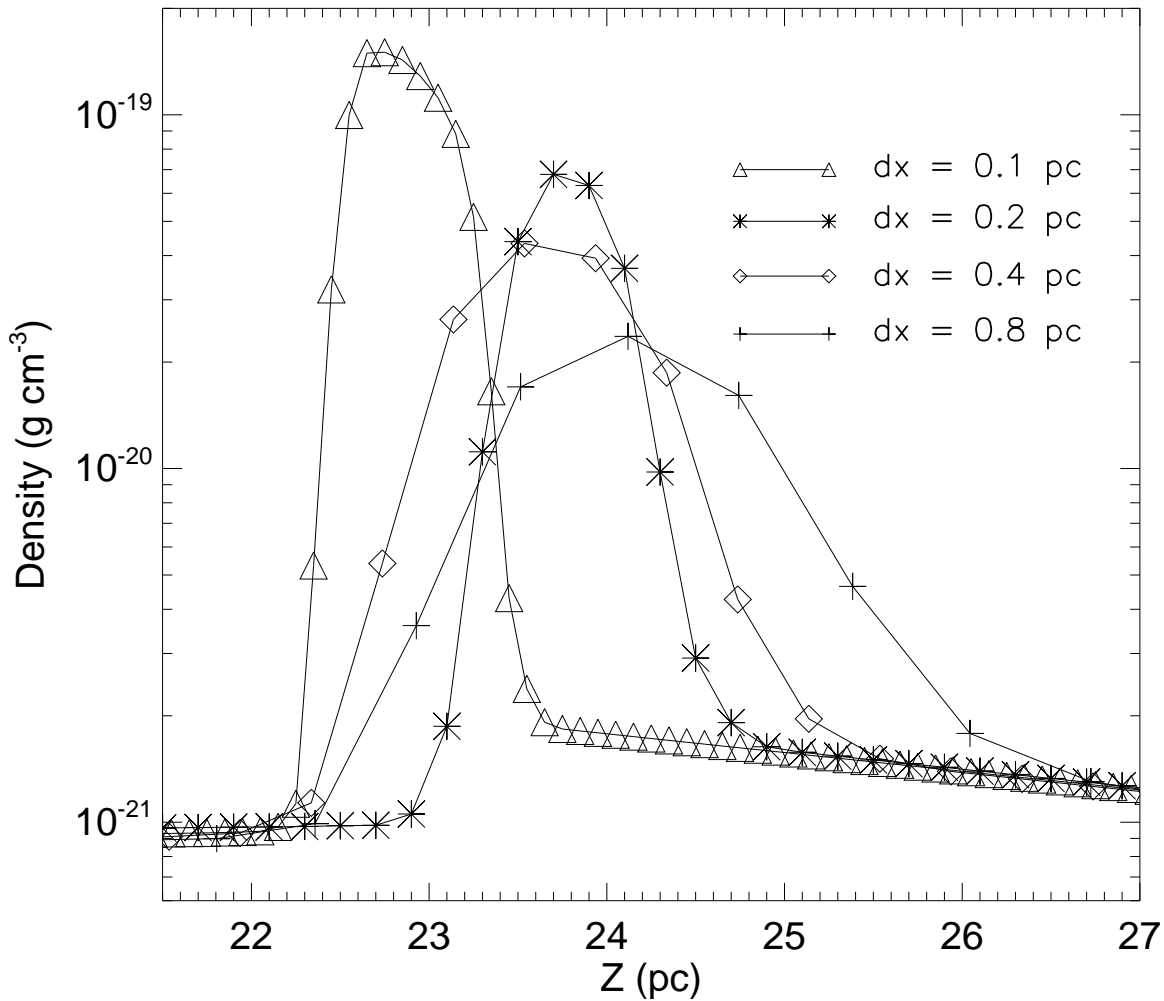} 
\figcaption{ Density
profiles at the outer shock fronts in the vertical direction at
$t=0.06$ Myr for 0.1~pc resolution ({\em triangles}; model X1-0), and at
$t=0.05$ Myr for 0.2~pc  ({\em stars}; X1), 0.4~pc ({\em
diamonds}; X1-2), and 0.8~pc ({\em crosses}; X1-4) resolution. The shell is better
resolved with a higher resolution, but still not fully resolved even
at 0.1 pc resolution.  (Note that the size of the highest resolution
bubble is slightly smaller only because of the smaller
source region.)
\label{dres}
}
\end{figure}

We show the density distributions of our fiducial model 
with different cooling curve cut-offs of $T_{floor}=10^2$ K (U1) and $10^4$ K (X1) at 
the time of blowout in Figure~\ref{L43}.  
Figure~\ref{L43} shows that the temperature floor
we choose for our cooling function appears to have a negligible influence on the
evolution of bubbles in our models.  

However, closer examination reveals that the density of shocked shells
is approximately the same in both simulations despite the difference
in temperature floor. For example, at $t=0.05$ Myr before any
fragmentation occurs, the shell density in both simulations is
$\sim8\times10^{-20}\mbox {g cm}^{-3}$ in the vertical direction where
the background disk density is $\rho_{bg}=2\times10^{-21}\mbox{ g
cm}^{-3}$.  This is because the density peak in the simulations is
limited by resolution in these models, not by the strength of cooling.
The shock velocity in the vertical direction at $t=0.05$~Myr is
$\sim230\mbox{ km s}^{-1}$, or Mach number ${\cal M} = 15$ in $10^4$~K
background gas. The immediate post-shock temperature is then $T =
7.6\times10^5$ K. This shocked gas quickly cools to or below $10^{4}$
K because the exponential cooling time (e.g. Mac Low \& McCray 1988)
is very short,
\begin{equation} 
\label{cool}
\tau_{cool} = 3kT/4n_{bg}\Lambda \approx64\mbox{ yr},
\end{equation} 
with the mean mass per particle $\mu = 14/22~m_{H}$
for ionized gas, and $\Lambda(T)=4.1\times10^{-23}
\mbox{ erg cm}^3\mbox{ s}^{-1}$ from the MacDonald \& Bailey (1981) cooling curve. 

The shell density expected from an isothermal shock will then be
$\rho_{shell}=\rho_{bg} {\cal M}^2\approx 5\times10^{-19}\mbox{ g
cm}^{-3}$ with Mach number ${\cal M}=15$ if $T_{floor}= 10^4$ K.  If
$T_{floor} = 10^2$~K, the shell density will reach values even greater
than the isothermal value.  However, since the shell density is far
from being resolved even with 0.1 pc resolution in our simulations,
the cooling floor has a negligible influence on our models.

\subsubsection{Effects of Source Region on Shell Fragmentation}
The top left panel of Figure~\ref{L421} shows the density distribution
of our fiducial model with $T_{floor}=10^4$ K, but with a source
region of a smaller radius of 5 pc or 25 zones (S1). The density
structure of S1 looks very similar to that of X1, but the
fragmentation by R-T instability is slightly suppressed. This is
because a smaller number of cells is covering the edge of the
spherical source region in which the density is imperfect.  This
imperfection creates a perturbation which gets amplified by
hydrodynamic instabilities.  A much smaller source region will make a
bigger difference in the amount of fragmentation.  It is important to
note that we are anyway probably underestimating the degree of
fragmentation since our bubble sweeps up a smooth, one-phase ISM
instead of real ISM with density fluctuating strongly from the mean
(see Cooper et al.\ 2008), and we have only a small, central source
region instead of an extended distribution of supernovae (cp.\ Fragile
et al.\ 2004).

\section{Comparison to Observations}
\label{sec:observation}
In this section, we first use the ballistic approximation to justify
making a comparison between the observations and our simulations in
which the bubbles are evolved only up $t\ll1$ Myr.  Then, we identify
gas parcels likely to produce Na~{\sc i} absorption, and simulate
observations of this gas along sightlines towards the galactic
nucleus. The velocity spread, the mass-weighted velocity, and the
maximum velocity are compared for different viewing angles.  The
velocity gradient across one of these winds is also studied for
comparison to longslit spectra. We then compare our models to the
observed Na~{\sc i} absorption profiles.

\subsection{Ballistic Approximation}
\label{subsec:ballistic}

We use a limited grid size in order to maintain high resolution, so we
simulate the evolution of bubbles only up to the time of blowout.  We
showed in the previous section that the bubbles blow out very early,
at $t\ll1$ Myr, because of the small scale height of our molecular
disk.  For comparison, ultraluminous starbursts have ages of up to
50~Myr (Murphy et al.\ 2001).  To extrapolate our computational
results to later times, we use the ballistic approximation (Zahnle \&
Mac Low 1995; Fujita et al.\ 2004) that after blowout, shell fragments
travel on radial ballistic orbits in the gravitational potential of
the galaxy, with no further accelerations by gas pressure gradients.

The strength of this approximation depends on the behavior of the
wind, and conditions in the region into which the wind
penetrates. This approximation was successfully used by Zahnle \& Mac
Low (1995) to follow the ejecta of a typical Shoemaker-Levy 9 fragment
falling back to Jupiter's atmosphere, and was found to give results
quantitatively consistent with the observations of the size and shape
of the infrared bright spots.  That case does differ from the
starburst case in having neither an ongoing wind, nor a complex
density distribution above the site of the explosion.  In the
starburst case, continuing supernova explosions at late times drive an
ongoing wind (see Figure~\ref{lmech}), while the mergers that drive
most starbursts yield a complex geometry above the site of the
blowout.  However, even if the stronger winds expected from supernovas
at late times in an instantaneous starburst do accelerate the
fragments further, the ballistic approximation will still yield a {\em
  lower} limit to their velocities.  Since the question of most
interest is whether cold gas can be accelerated to such high
velocities, a lower limit is already useful.  The complex geometry of
tidal tails may also not be of great concern, as models suggest that
they only cover a small fraction of the solid angle visible from the
nucleus, and so are unlikely to be dynamically dominant.

Under the ballistic
approximation, the equation of motion for a shell fragment at a
distance $r$ from the galactic nucleus is
\begin{equation}
v(r) = \left\{v^2(r_b)+2[\Phi(r_b)-\Phi(r)]\right\}^{1/2},
\label{ba}
\end{equation}
where $\Phi(r)$ is the total halo and disk potential, and $v(r_b)=v_b$
is the shell velocity at blowout at a position of $r_b$.

\begin{figure}
\includegraphics[width=\textwidth]{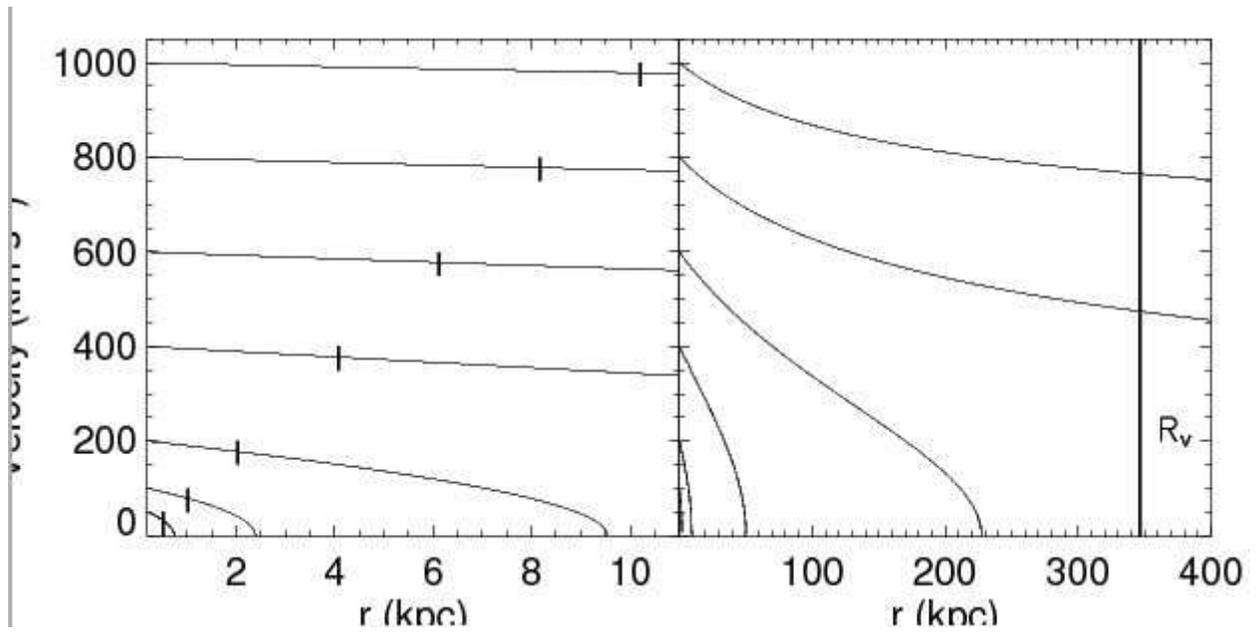} 
\figcaption{The velocity
predicted by the ballistic approximation for shell fragments as they
expand radially in the halo from $r_b=0.2$ kpc with initial (blowout)
velocities, $v_b=50$, 100, 200, 400, 600, 800, and 1000 km s$^{-1}$.
The {\em left} panel shows the behavior near the galaxy, while the
{\em right} panel captures the full extent of the halo.  Radii at
$t=10$ Myr ({\em left panel}) and the virial radius ({\em right
panel}) are noted with {\em thick lines}.
\label{ballistic}
}
\end{figure}

In Figure~\ref{ballistic}, we plot $v(r)$ for several initial
velocities starting at a fixed initial position, $r_b=200$ pc, just
above our model disk. By setting $r_b=200$ pc, we can ignore the disk
potential which is negligible above the disk compared to the halo
potential.  Thus we set $\Phi=\Phi_{halo}$ in equation~(\ref{ba}) and
solve the equation analytically using equation~(\ref{nfw}).  The upper
limit to the radius reached by a shell fragment at time $t$ is given
by the linear approximation $r \le v_bt$, because the shell will only
decelerate in the potential.  We show actual radii for different
initial velocities at $t=10$ Myr by vertical ticks in the left panel
of Figure~\ref{ballistic}.  Although Figure~\ref{ballistic} is plotted
as a function of radial velocity, we find similar results if total
velocity is used instead, because shell fragments are traveling in
nearly radial directions.  This figure shows that the linear
approximation is quite good for $v_b > 100 \mbox{ km s}^{-1}$, so we
use it to follow the cold gas for times of order 10~Myr.

For times $t\gg100$ Myr, the linear approximation fails and most gas
falls back to the center after reaching heights of a few hundred kpc.
Only gas with $v_b\ga1000\mbox{ km s}^{-1}$ can completely escape the
potential of the disk and halo.  (The halo alone has an escape
velocity $v_{esc}=800\mbox{ km s}^{-1}$ at radii less than $0.01
R_v$.)  We show below that very little gas actually escapes the halo,
with $\la0.1\%$ of the total shell mass accelerated above $v_b
\ga1000\mbox{ km s}^{-1}$ in our fiducial model.  Our results confirm
the result from previous simulations of smaller galaxies that the loss
of ISM mass is inefficient (e.g. Mac Low \& Ferrara 1999; D'Ercole \&
Brighenti 1999).  However, significant mass is circulated
over scales of 10 kpc, presenting a significant absorption cross
section as suggested by Martin (2006).

\subsection{Cool Gas}
\label{subsec:coolgas}

\begin{figure}
\includegraphics[width=\textwidth]{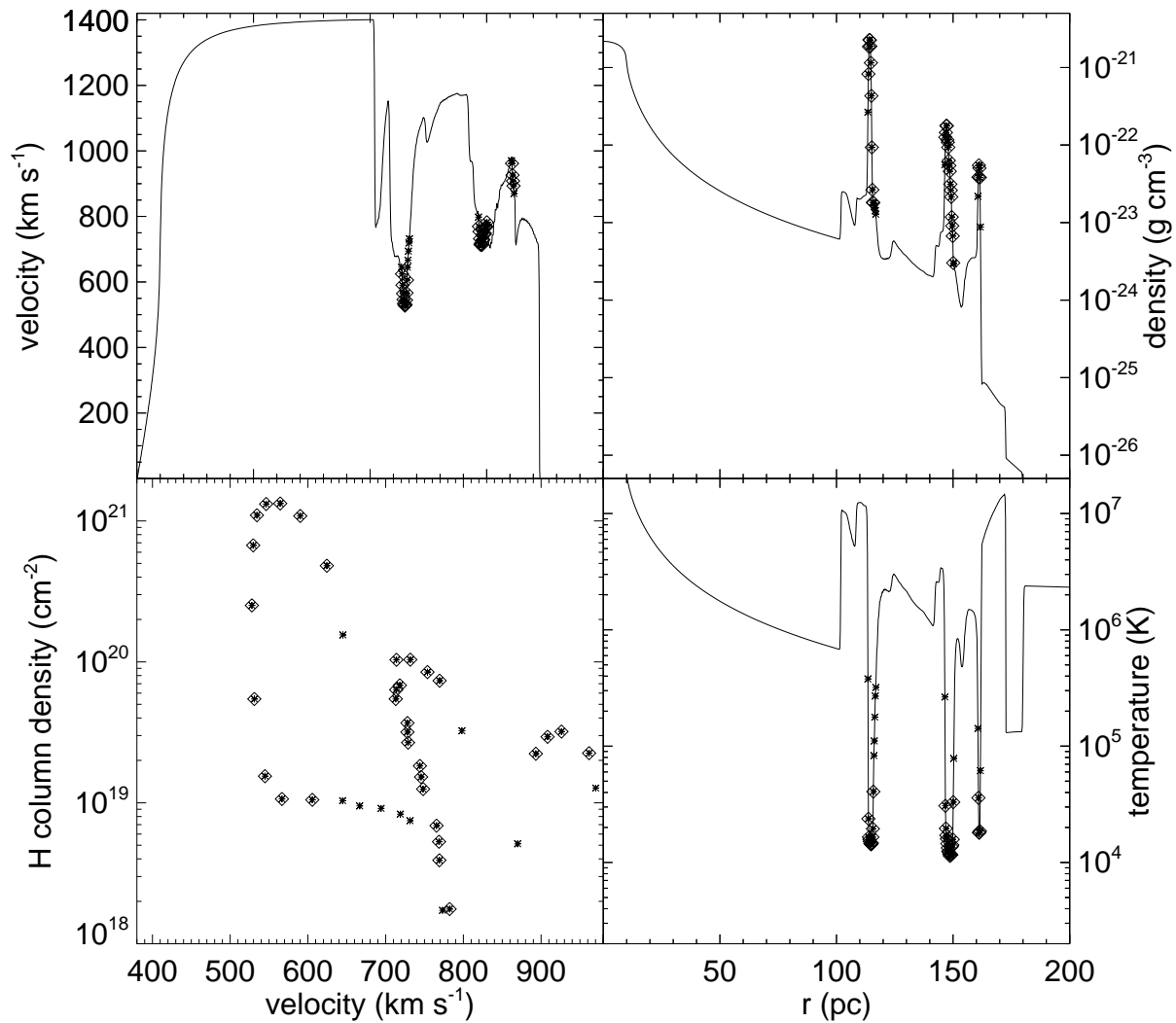} 
\figcaption{  In the {\em bottom left} panel, the
distribution of column density for the cold gas as a function of
radial velocity is plotted. The other three panels plot radial
profiles of radial
velocity ({\em top left}),  density ({\em top right}), and 
temperature ({\em bottom right}) along a line of
sight through the center at an angle of 19$^{\circ}$ from the vertical
axis in model X1 at the end of the simulation.  The radial profiles all use the
horizontal axis labeled on the bottom right. Regions of cold gas
with $T<5\times10^4$ K (which we take to be Na~{\sc i} absorbing gas)
are shown in {\em diamonds}. 
\label{line}
}
\end{figure}

Only the coolest gas can be observed in Na~{\sc i} absorption.  We
chose a temperature cut-off of $T<5\times10^4$ K to trace this gas
because cooling remains numerically limited even with our highest
resolution grid.  Gas below this cut-off is so dense that the cooling
time (eq.~\ref{cool}) is very short, $\tau_{cool}< 0.01$ Myr, so it is
physically expected to reach temperatures where Na~{\sc i} is present.
As an example, the least dense shell fragment in Figure~\ref{line} has
$\rho=3.0\times10^{-24}\mbox{ g cm}^{-3}$ and $T=3.3\times10^4$~K.
Equation~(\ref{cool}) gives an exponential cooling time
$\tau_{cool}\approx 0.007$ Myr, using
$\Lambda(T)\approx4.2\times10^{-23}\mbox{ erg cm}^3\mbox{ s}^{-1}$
from the MacDonald \& Bailey (1981) cooling curve.

\begin{figure}
\includegraphics[width=\textwidth]{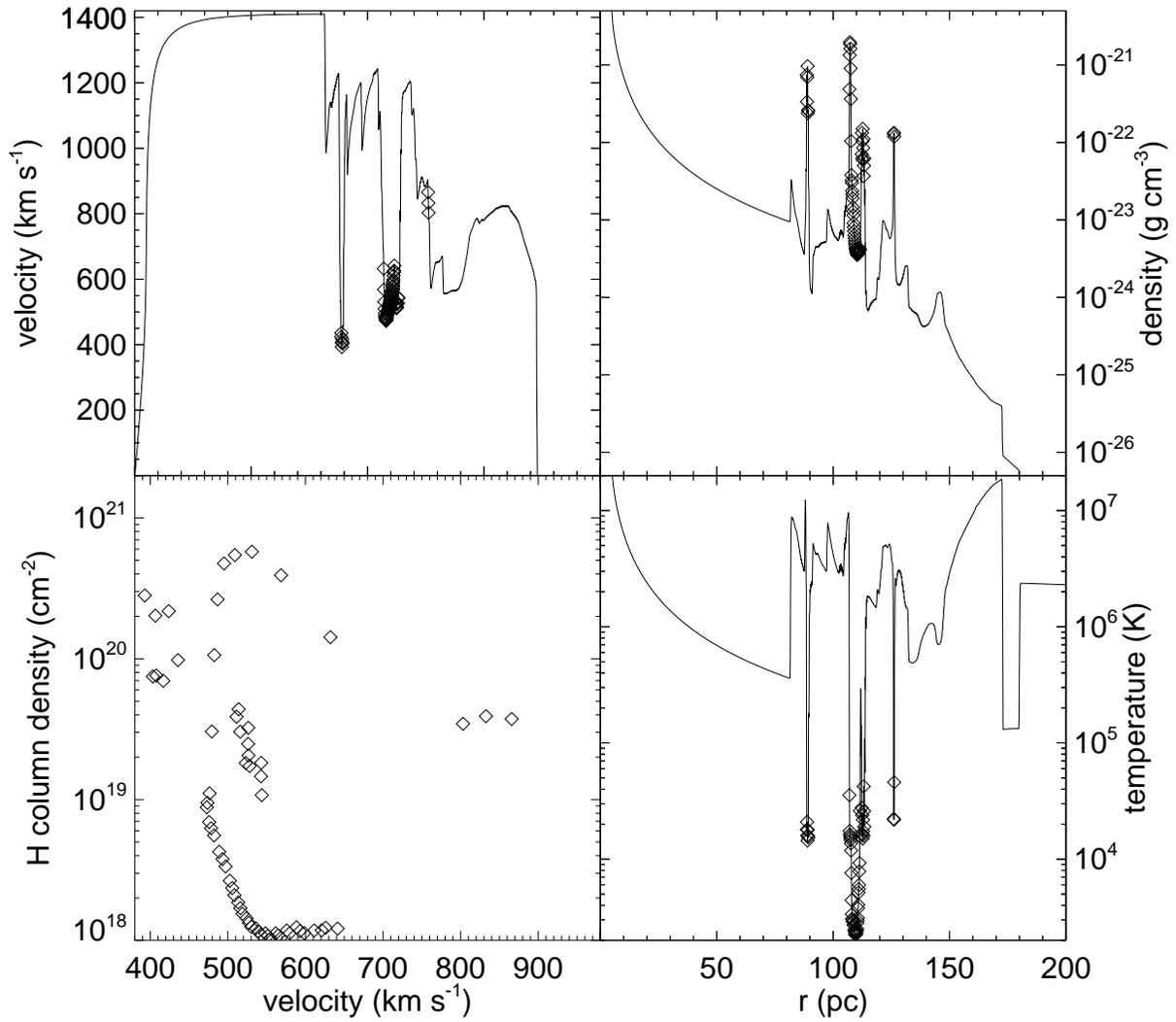}
\figcaption{The same as in Figure~\ref{line} for the same standard
  simulation, but with our highest resolution of 0.1 pc (X1-0).
  Again, the three radial profiles ({\em top right, top left, bottom
    right}) all use the radial axis given on the bottom right.
\label{bline}
}
\end{figure}

Figures~\ref{line} and~\ref{bline} show the gas density, temperature,
and radial velocity in models X1 and X1-0 along a line of sight
through the central continuum source at an angle $\theta=19^{\circ}$
from the vertical axis.  Notice that our temperature cut-off picks up
only the densest and the coolest parts of the shells excluding
numerically diffused interfaces between the shells and the hot gas. 

We compare the characteristics of this temperature-selected gas to the
measured properties of Na~{\sc i} absorption in ultraluminous
starbursts.  As argued above in \S~\ref{subsec:sf}, our assumption of
a central point source of wind luminosity likely represents a lower
limit to the amount of cold gas entrained.  Fragile et al.\ (2004)
found that, in galaxies where a central energy source ejects nearly
all its kinetic energy in the hot gas (Mac Low \& Ferrara 1999),
distributed supernovas deposit as much as half of the energy in the
cold gas.  However, much of this energy is radiated promptly rather
than converted to kinetic energy, so determining the significance of
distributed supernovas requires further work.

\subsubsection{Cool Gas Column Densities}

\begin{figure}
\includegraphics[width=\textwidth]{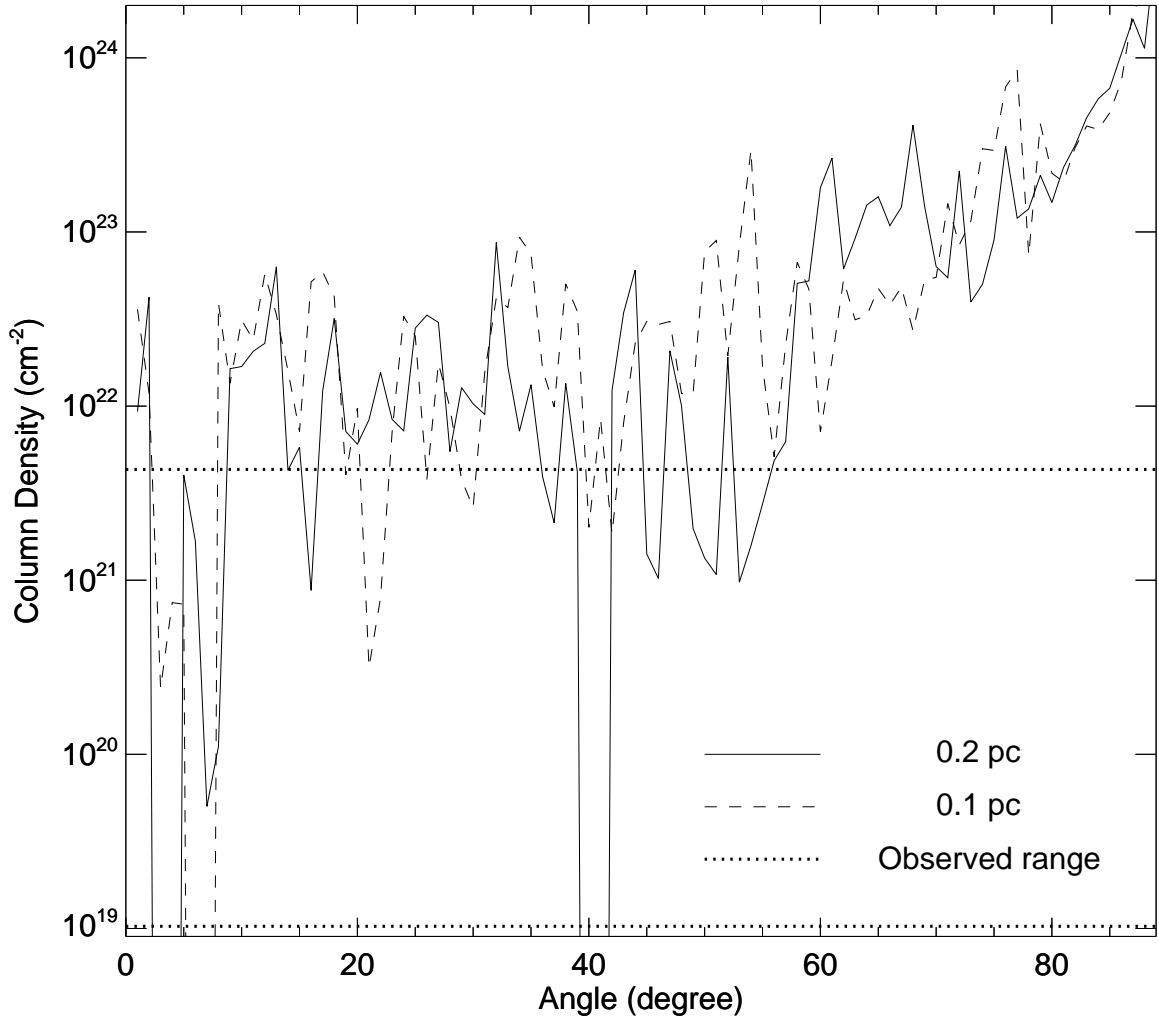}
\figcaption{The column density distributions of cool gas at sightlines
through the center as a function of angle extended from the vertical
axis in models with 0.2~pc ({\em solid line}; model X1) and 0.1~pc ({\em
dashed line}; X1-0) resolution. The observed range of column density
inferred from observations of Na~{\sc i} absorption profiles is
$1.0\times10^{19}-4.3\times10^{21}\mbox{ cm}^{-2}$, shown in {\em
dotted lines}. Note that the column densities from the models will be reduced
over time due to spherical expansion.
\label{hcolumn}
}
\end{figure}

Figure~\ref{res} shows that the lines of sight plotted in
Figures~\ref{line} and \ref{bline} go through multiple distinct
shells, adding to a total column density of cool gas $N_H
\sim7\times10^{21}\mbox{ cm}^{-2}$.  The inferred column densities in
ultraluminous starbursts are a bit lower, exceeding
$10^{21}$~cm$^{-2}$ in only one out of four ultraluminous starbursts
(Martin 2005).  Since the model column densities fluctuate by as much
as an order of magnitude between neighboring sightlines, we plot the
column density distribution as a function of angle in
Figure~\ref{hcolumn}.  We find that most sightlines with $N_H > 5
\times 10^{22}$~cm$^{-2}$ lie within 30$^\circ$ of the galactic disk
where the shells are usually not subject to hydrodynamic
instabilities. The typical value is $N_H \sim 10^{22}$~cm$^{-2}$,
similar to the largest column density estimated from observations of
ultraluminous starbursts (Martin 2005, 2006; Rupke et al. 2002).

\subsubsection{Geometric Dilution}

The column densities found in our model are measured at a time very
early in the lifetime of the wind, when it has not expanded far away
from the galaxy. However, these winds will typically be observed at
later times, when the wind has expanded further.  Geometric dilution
will then reduce the column density along any particular line of sight even
if the gas simply expands radially outwards on ballistic orbits.  
For spherical geometry (of any opening angle) and constant velocity
flow, the volume-averaged gas density declines with increasing radius.
The amount of dilution is therefore determined by the radial
advance of the innermost cold gas. At early times, see Figure~\ref{res}
this innermost cold gas lies at a radius of $\sim 100$~pc, near the wind
termination  shock.  Assuming the cold gas directly above the
starburst region flows upward on a ballistic path, we would expect
it to move outward by a factor of roughly 100 in the next 25~Myr,
reducing the cold gas column density by the same factor.

\subsubsection{Resolution Effects}

A second reason we overestimate the cool gas column density is that
the fragmentation of the shells is limited by numerical resolution, as
well as our assumption of azimuthal symmetry.  However, as discussed
above in \S~\ref{subsec:effects-res}, the overall kinematics described
in \S~\ref{subsec:ballistic} should remain similar unless shell
fragments are entirely mixed with hot gas and destroyed.  To
distinguish real physical effects from numerical artifacts, we study
the kinematics of the cold gas along a $\theta=19^{\circ}$ line of
sight in our fiducial simulation X1 and the high-resolution simulation
X1-0.

The first two fragments from the center are
parts of a filament remaining from the initial fragmentation of the swept-up
supershell due to R-T instability  
at $t\sim0.1$ Myr. These high-density peaks are traveling with similar
velocities $v\sim400-500\mbox{ km s}^{-1}$ in both simulations
and dominate the column densities of cool gas in the studied sightlines. 
The outermost shell in X1 keeps sweeping up high-altitude disk gas, 
while in X1-0 this shell further fragments by R-T instability to the last two fragments
in Figure~\ref{bline}. The coherence of the shell in X1 clearly occurs
because of the lower resolution in that run.
However, the shell in X1 and the outermost fragment of the two in X1-0 travel 
with similarly high velocity of $\sim800-900\mbox{ km s}^{-1}$, though they 
contribute very little to the total column density. 

The hot wind overruns and shocks the first fragment in X1 and the
first two fragments in X1-0.  As a result, secondary R-T and K-H
instabilities act on them, removing gas from the cold cloud and mixing
it into the hot wind.  In both cases, these clumps of gas are resolved
by about $\sim10$ cells, demonstrating that this is a minimum size
below which fragments tend to survive because the secondary
instabilities cannot be resolved.  Mac Low \& Zahnle (1994) show that
fragments must be resolved by at least 25 zones to resolve the
secondary instabilities.  Fragments smaller than that remain as
artificially massive clouds in our simulations, overpredicting the
cool gas column densities.

Klein et al.\ (1994) suggest that these clumps of gas should be
destroyed, via hydrodynamic instabilities, over $\sim10$ shock
crossing timescales
\begin{equation}
t_{cc} = r_c(v_w-v_c)\sqrt{\frac{\rho_{c}}{\rho_{w}}}
=(0.04\mbox{ Myr})
\left(\frac{r_c}{2\mbox{ pc}}\right) 
\left(\frac{v_w-v_c}{500\mbox{ km s}^{-1}}\right)
\left(\frac{\rho_c/\rho_w}{100}\right)^{1/2},
\end{equation}
where $r_c$ is the radius of the clump, $\rho_w$ and $\rho_c$ are the
density, and $v_w$ and $v_c$ are the velocity of the wind and the
clump, and the scaling parameters hold for typical 10 zone clumps in
our model.  In reality, $t_{cc}$ may be substantially longer because
the density ratios $\rho_c/\rho_w$ are probably underestimated: shell
densities, and thus clump densities, are underresolved, while the wind
density may be overestimated by our mass-loading scheme.  We need
$\rho_c/\rho_w\approx10^{4}$ for these clumps to survive for more than
$\sim5$ Myr, which may just be reachable in a rapidly diverging wind.

Moreover, recent numerical studies clumps may be stabilized against
Kelvin-Helmholtz and R-T instabilities, reducing mass loss, by either
thermal conduction (Marcolini et al.\ 2005, Vieser \& Hensler 2007),
or even weak magnetic fields (Mac Low et al.\ 1994; Shin et al.\
2008).  Since we are not able to address the fate of shell fragments
further in this study, we assume the bulk of the cool mass remains
cool for the duration of starburst to be observed as Na~{\sc i}
absorbing gas.  Larger, denser clumps are more likely to survive, in
reality.

\subsection{Velocity Spread}
\label{subsec:vspread}

The bottom left panels of Figure~\ref{line} and \ref{bline} 
show the distribution of column density of cool gas $N_H$ as a function of 
its radial velocity $v_r$ 
at the end of the simulations. So long as the linear version of the
ballistic approximation is correct, the velocity spread remains constant.
Cool gas is seen over a wide range of velocities $>450\mbox{ km
  s}^{-1}$ at both resolutions. Shells that fragment earlier have been
accelerated less, and so their fragments travel more slowly.  

\begin{figure}
\includegraphics[width=\textwidth]{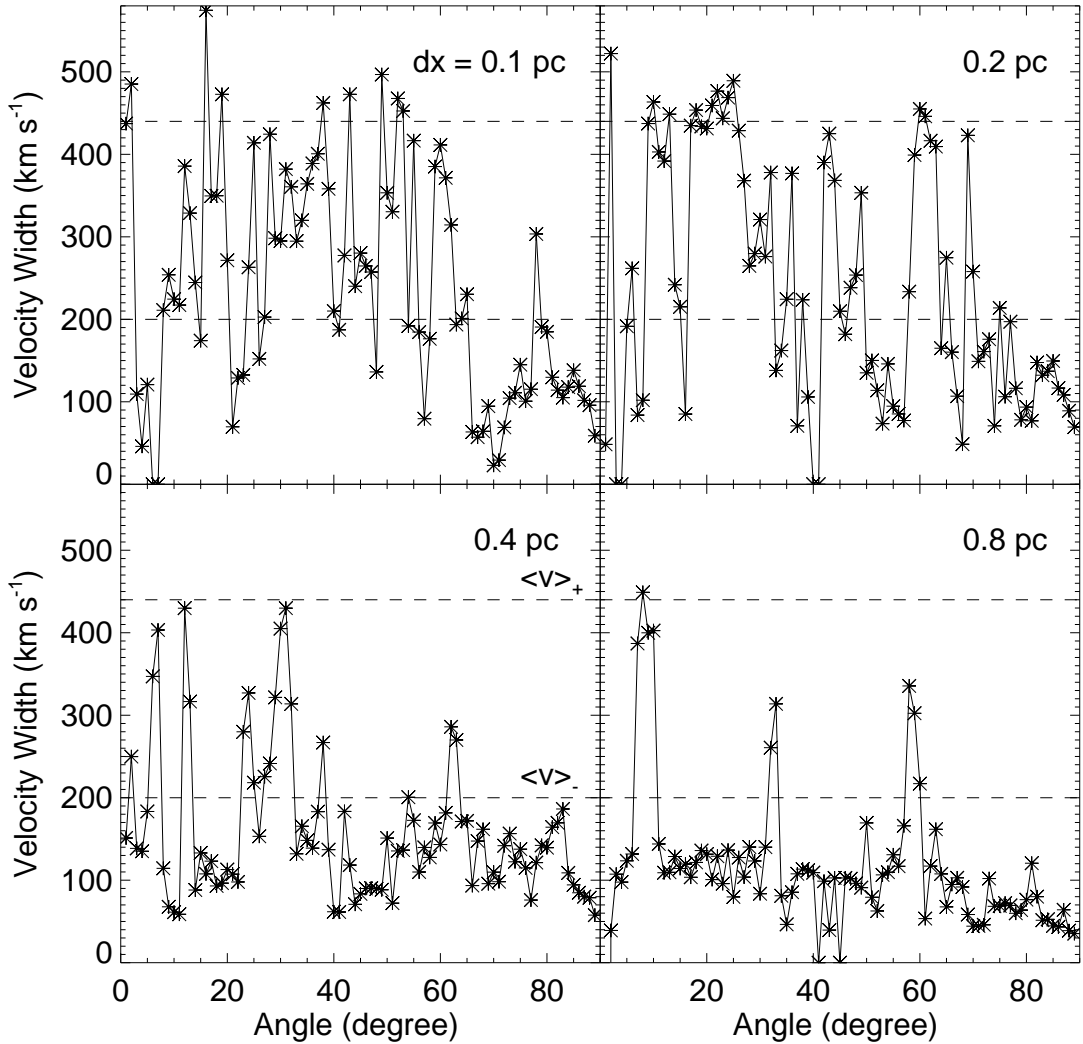}
\figcaption{The velocity widths are shown as a function of angle from 
the vertical axis at every degree for models with grid resolutions of
$dx = 0.1$, 0.2, 0.4, and 0.8~pc (models X1-0, X1, X1-2, and X1-4). Note that the
highest two resolutions appear to display converged behavior.
The upper and lower limits of the observed average line width in Na~{\sc i} 
absorption are shown in ({\em dashed lines}): $\langle v \rangle=320\pm120\mbox{ km s}^{-1}$.
\label{lw}
}
\end{figure}

We now explore whether absorption from multiple shell fragments can
explain the large absorption line widths measured for ultraluminous starbursts by
considering the effects of resolution and viewing angle.
Figure~\ref{lw} shows the distribution of velocity widths $\Delta v$
seen in cool gas with $T<5\times10^4$ K in runs X1-0, X1, X1-2, and
X1-4, with grid size increasing from 0.1 to 0.8~pc, along sightlines
spaced by $1^{\circ}$ from the vertical axis.  We define $\Delta v$ as
the difference between the maximum and the minimum velocities seen in
the cool gas along a given sightline---that is, full width at zero.
This resolution study demonstrates that the fraction of sightlines
with large velocity spread increases as the resolution improves up to
0.2 pc, but then appears to converge.

Comparison to Figure~\ref{res} shows that the largest velocity widths
are seen in sightlines intersecting the largest number of fragmented
shells.  Better resolving post-shock densities in the swept-up,
cooled, shells leads to quicker shell fragmentation and reformation,
producing a larger number of shells, and thus the wider range of
velocities in the fragments.

Figure~\ref{lw} also shows that most sightlines in X1 and X1-0 are above 
the average observed line width $\langle v \rangle=320\pm120\mbox{ km s}^{-1}$ ({\em dashed lines}). 
Observers measure the average of the sightlines toward continuum sources subtending
about an arcsecond, and this corresponds to a length of
  $1.84~h_{0.7}\mbox{ kpc}$ at redshift 0.1. 
Assuming that the observers are likely to view ULIRGs at $t=5-10$ Myr
after the onset of starburst, we can suggest all the shell fragments within
$\sim20^{\circ}-40^{\circ}$ at a given sightline in our simulations will
contribute to the absorption profiles.
Hence, our models suggest that 
observers will measure a large line width regardless of viewing angle.

To quantify this, we compute the average velocity width of all sightlines 
in an axisymmetric cone within $\theta$ as 
\begin{equation}
\overline{\Delta v}(\le\theta) = \frac {\int_0^{\theta} \Delta
  v(\alpha) \sin(\alpha) d\alpha} {\int_0^{\theta} \sin(\alpha)
  d\alpha}. 
\label{av}
\end{equation}

The value of $\overline{\Delta v}$ is lowered by sightlines near the
disk midplane where the bubbles are not blowing out and sightlines
intersecting holes made by blowout with nearly zero $\Delta v$. The
average velocity widths are clearly smaller for lower resolution runs,
however. We find $\overline{\Delta v}(\theta\le30^{\circ})$ is
220~km~s$^{-1}$ for run X1-0, and 210~km~s$^{-1}$ for X1, but only
110~km~s$^{-1}$ for X1-4. 

\subsection{Average Velocity}
\label{subsec:AMvelocity}

\begin{figure}
\includegraphics[width=\textwidth]{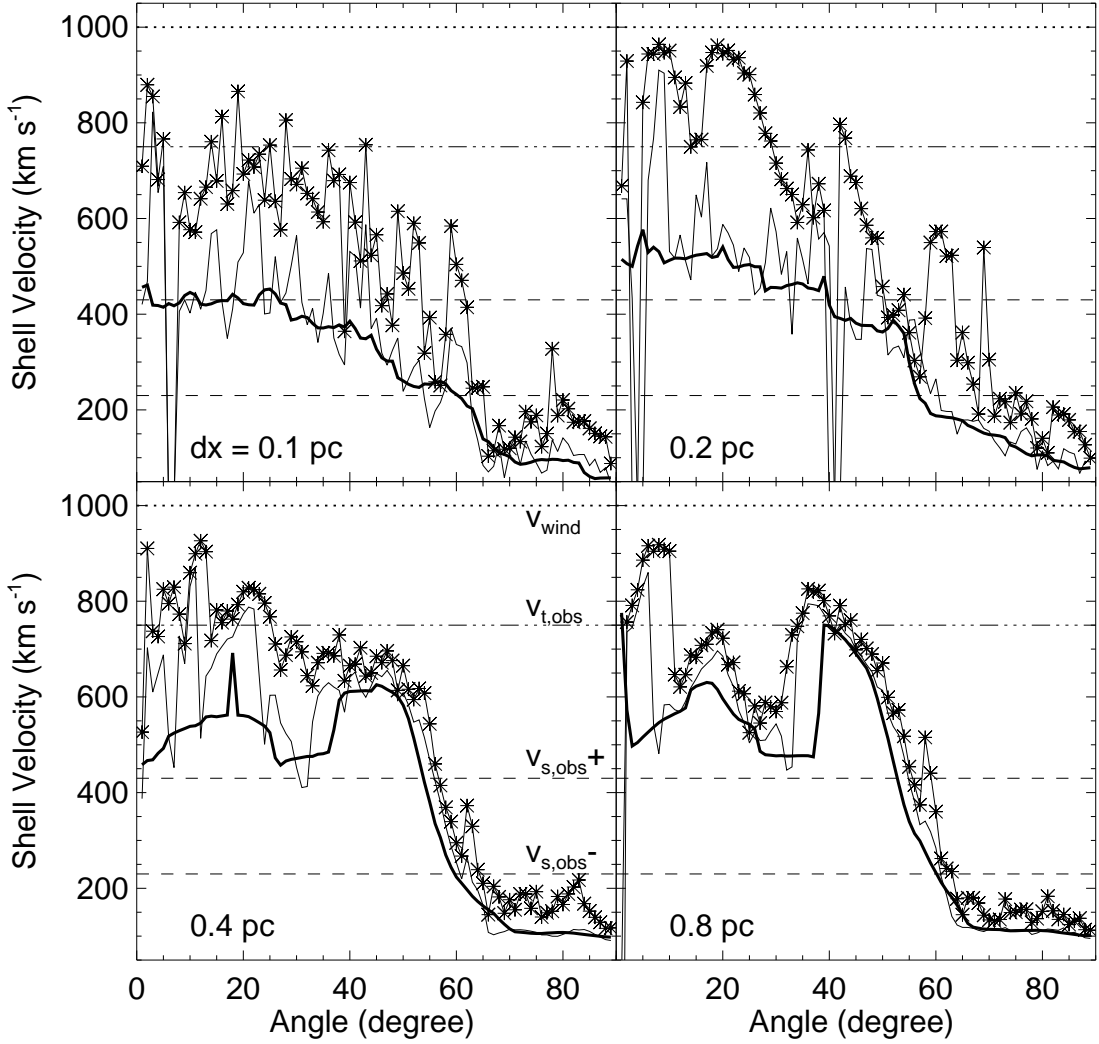} \figcaption{The terminal
velocity ({\em solid line with asterisks}) and the average
mass-weighted velocity ({\em thin solid line}) of cool gas are plotted
as a function of angle from the vertical axis along sightlines through
the galactic center in models with grid resolutions of $dx = 0.1$,
0.2, 0.4, and 0.8~pc (models X1-0, X1, X1-2, and X1-4).  Since shell
mass varies substantially with angle, we also plot the mass-weighted
average velocity within a $10^{\circ}$ arc $v_{av,10}$ at each angle
({\em thick solid line}).  We show the terminal velocity of the low
density wind $v_{wind}\approx1000\mbox{ km s}^{-1}$ in {\em dotted
line}, the observed average shell velocity $v_{s,obs}=330\pm100\mbox{
km s}^{-1}$ in {\em dashed line}, and the observed average terminal
velocity $v_{t,obs}=750\mbox{ km s}^{-1}$ in {\em dash-dot-dot line}.
\label{term}
}
\end{figure}

Figure~\ref{term} shows the distribution as a function of angle with
degree spacing of mass-weighted average velocities of cool gas
$v_{av}$ ({\em solid lines}) for our standard resolution study (runs
X1-0 through X1-4).  The average shell velocities plotted may be
misleading, since shell mass differs significantly at each sightline.
Therefore, we also plot a mass-weighted average shell velocity across
a $10^{\circ}$ arc centered on each sightline $v_{av,10}$.  This ought
to be the quantity most directly comparable to observations of the
average velocity.  

On sightlines with multiple fragments, the average mass-weighted
velocity reflects the velocity of the more massive fragments.  These
are fragments of the initial swept up shell, which follow ballistic
orbits with little acceleration after shell fragmentation. Thus,
average mass-weighted velocity tends to lie substantially below peak
velocity.

The observable quantity $v_{av,10}$ shows a clear converging trend
with resolution.  The converged value toward the pole appears to be
under 400~km~s$^{-1}$, with correspondingly lower values at other
angles, as shown in Figure~\ref{term}.  Figure~\ref{term} suggests
that cool gas in shells and their fragments will be observed to be
traveling with $v_{av,10}\approx200-350\mbox{ km s}^{-1}$ at angles to
the pole $\theta<60^{\circ}$, the angle where blowout occurs in all
models.

Our model is consistent with observations that constrain ULIRG wind
geometry.  Figure~\ref{term} shows that absorption at velocities $v >
100$~km~s$^{-1}$ is detected at all angles greater than $10^{\circ}$
from the disk plane.  It follows that 98 \% of all random, radial
sightlines would exhibit such outflows. This result is consistent with
15 of 18 observed objects showing such outflows in Martin (2005), and
a similar fraction seen by Rupke et al.\ (2005).

\begin{figure}
\includegraphics[width=\textwidth]{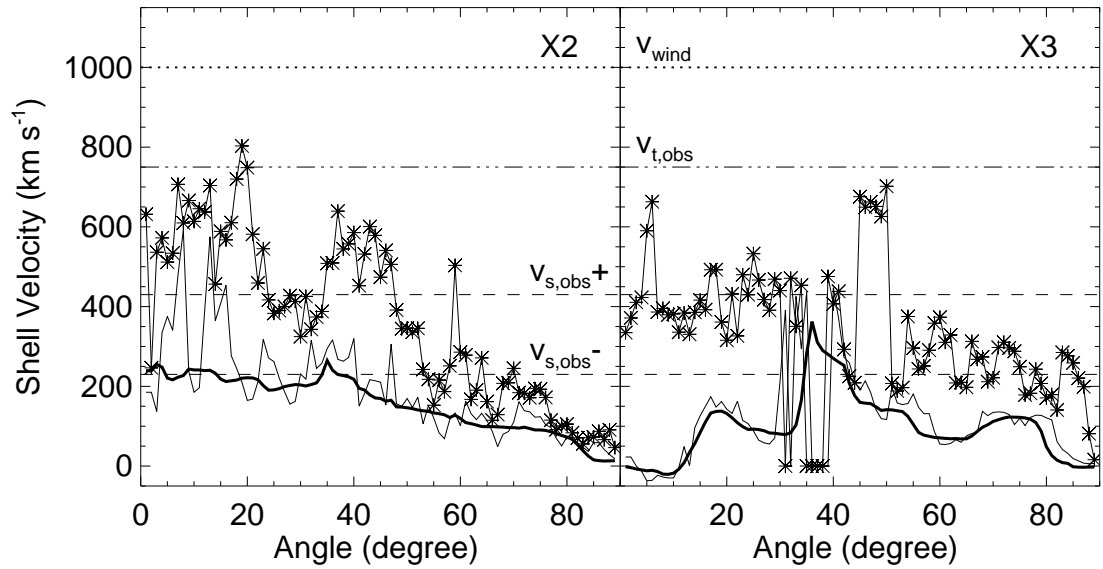}
\figcaption{Same as in Figure~\ref{term} for
  lower luminosity runs with $L_{mech} = 10^{41}$~erg~s$^{-1}$ 
  ({\em right}; model X3), and $10^{42}$~erg~s$^{-1}$ ({\em left}; X2). 
\label{lwt}
}
\end{figure}

Figure~\ref{lwt} shows the distributions of $v_{av}$ and $v_{av,10}$
for models X2 and X3 run with decreasing mechanical luminosity at the
same resolution as model X1.  The mass-weighted average velocities of
shells should depend on the mechanical luminosity $L_{mech}$ driving
the bubble by thermal pressure. In a spherical bubble, the dependence
would be $L_{mech}^{1/5}$. This dependence is actually slightly
stronger in our models of blowout. We find $v_{av,10}$ dropping from
400~km~s$^{-1}$ to $\sim 100$~km~s$^{-1}$ moving from model X1 to X3,
with each step having an order of magnitude lower mechanical
luminosity.  This gives an empirical relation closer to
$L_{mech}^{0.3}$.  Note the high spike in $v_{av,10}$ of X3 around
$\sim15^{\circ}$ is biased by a small amount of cool mass present in
its vicinity.  By way of comparison with observations, Figure 6 of
Martin (2005) shows velocities reaching $v = 700$~km~s$^{-1}$ for an
SFR of 1000~M$_{\odot}$~yr$^{-1}$. At an SFR of
1~M$_{\odot}$~yr$^{-1}$, the sparse data shown in that Figure suggest
$v = 30$--40~km~s$^{-1}$, while our empirical relation would suggest a
value $v \simeq 90$~km~s$^{-1}$.  More data at low SFRs and a broader
range of models will be required to establish whether there is truly a
discrepancy with the model results.

The mass-weighted average velocities $v_{av,10}$ in X1 and X2 agree
with the observed shell velocity, but $v_{av,10}$ in X3 is well below
the observed value.  In disks as massive as these, starbursts with
mechanical luminosity lower than $10^{42}\mbox{ erg s}^{-1}$ produce
the lower end of the outflow velocity range observed in ULIRGs, which
average $v_{s,obs}=330\pm100\mbox{ km s}^{-1}$ in the Martin (2005)
sample and $170\mbox{ km s}^{-1}$ in the column-density weighted
average velocities from the Rupke et al. (2005) sample.
Column-density weighting with velocity can be confounded by variation
in covering factor with velocity, as has been shown for AGN outflows
(e.g., Arav et al. 1999, 2005).  We cannot model covering factors well
with our two-dimensional simulations, so we postpone consideration of
the question of column-weighted velocity to future work. We do note
that overpredicting acceleration is much harder to do than
underpredicting it, though, so we think our model is likely to be
robust. The velocity spread, which is our most important result from
the simulations, is similar to that found in both ULIRG studies.

\subsection{Terminal Velocity}
\subsubsection{Resolution}
\label{subsub:term-res}

Figure~\ref{term} also shows the distribution of terminal velocity of
cool gas  $v_{term}$ at each sightline. We define the maximum
velocity at blowout as the terminal velocity a shell will ever aquire,
following the ballistic approximation
(\S~\ref{subsec:ballistic}). The maximum velocity of shell fragments is
very high, $v_{term}>500\mbox{ km  s}^{-1}$ at the angle $\theta<60^{\circ}$ where
blowout occurs in all the runs. This high-velocity cool gas is found 
in the fragments of the outermost shells in our highest resolution
model X1-0. The hot wind continues to accelerate a piece of shell,
sweeping up the ambient gas, until it fragments further.  As
discussed above, shell fragmentation is very sensitive to resolution.
Thus it is important to test the convergence of the mass of
high-velocity gas.

\begin{figure}
\includegraphics[width=0.5\textwidth]{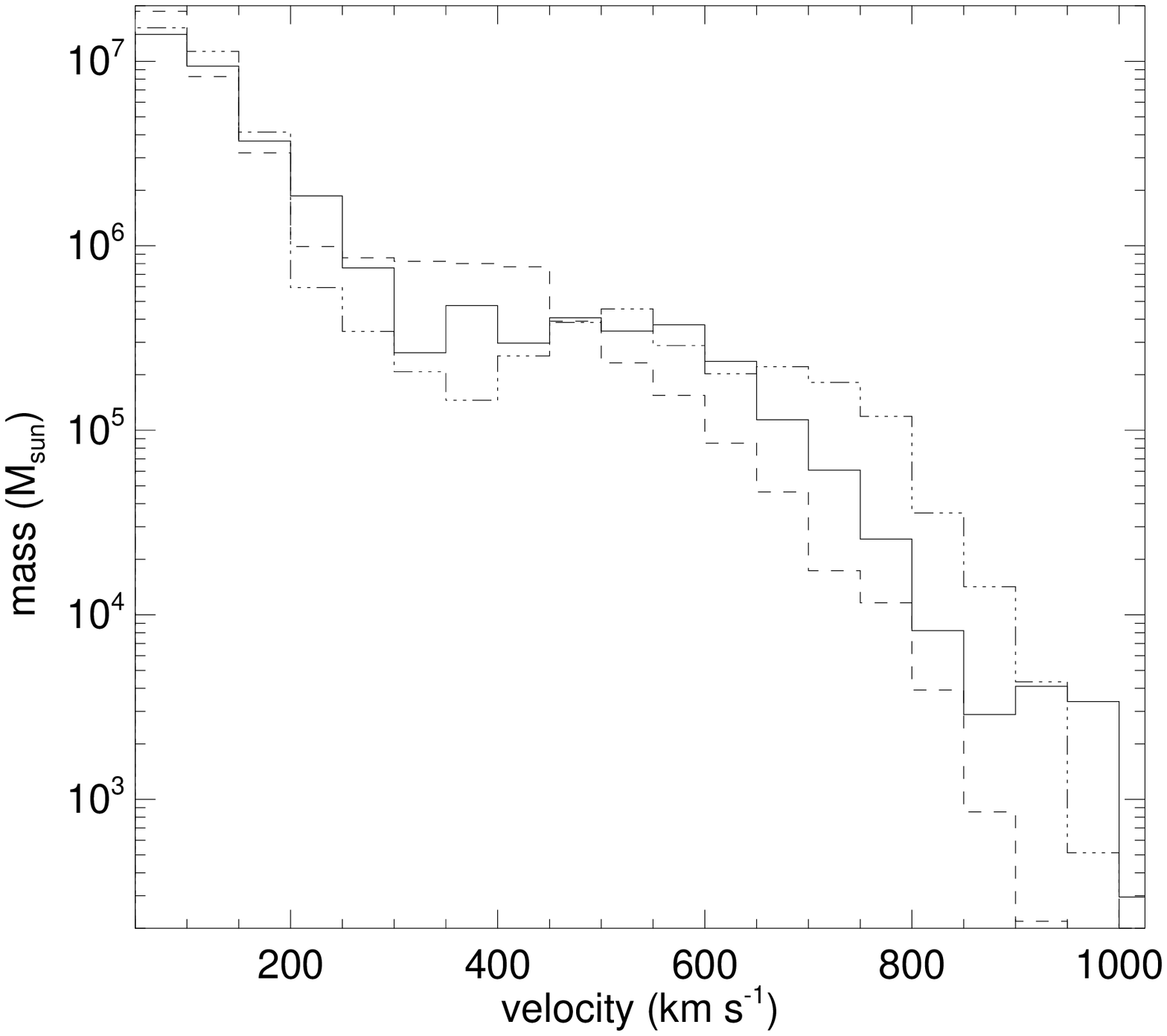}
\includegraphics[width=0.5\textwidth]{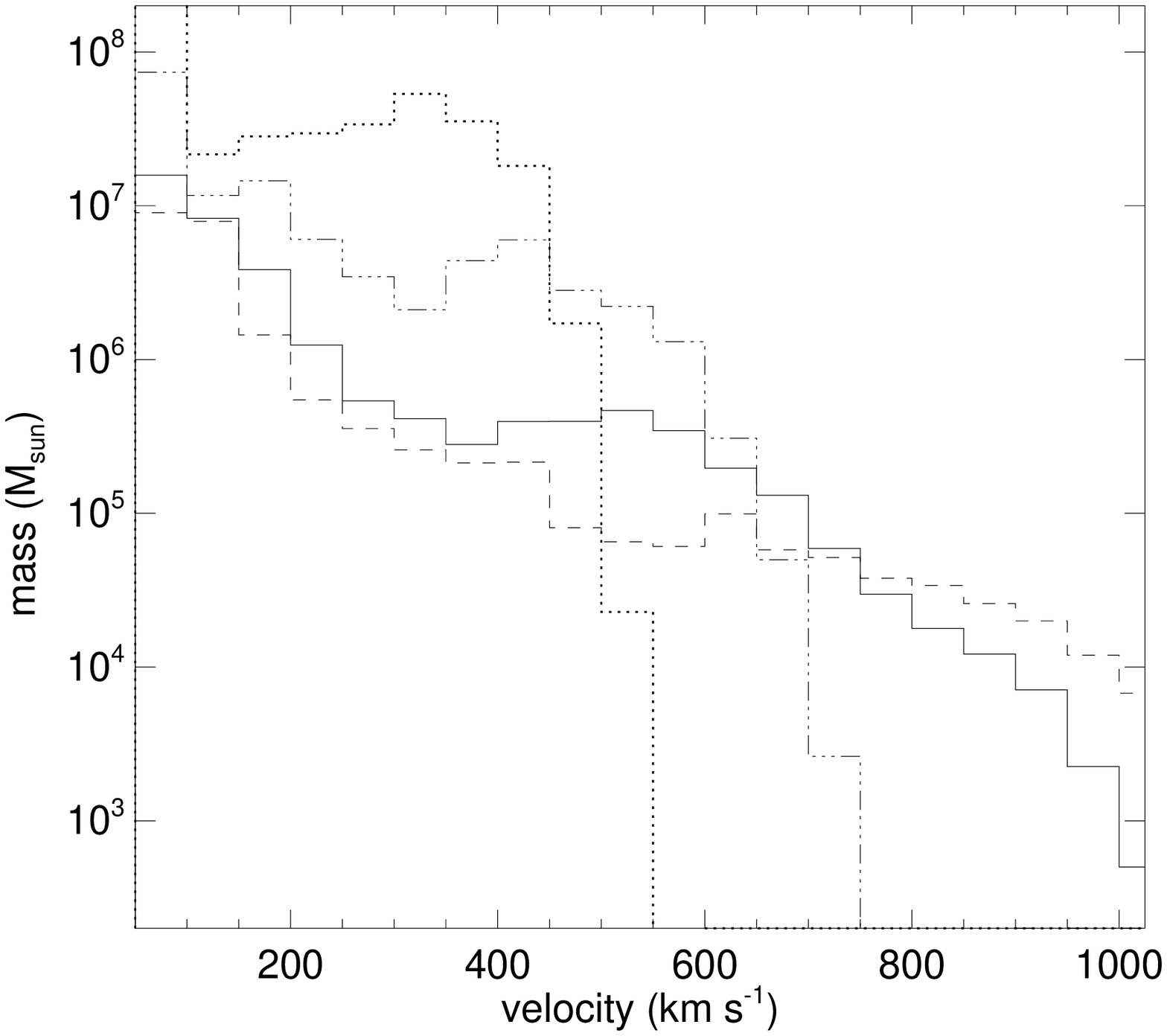} \figcaption{The mass
distributionsof all the cool shells and shell fragments as a function
of velocity for {\em (a)} models at $t=0.27$~Myr with increasing zone
sizes $dx = 0.1$~pc ({\em dashed line}; model X1-0), 0.2~pc ({\em
solid line}; X1), and 0.4 pc ({\em dashed-dot-dot line}; X1-2), and
{\em (b)} models at the blowout time ($t=0.22$, 0.27, 0.33, and 0.41
Myr) with increasing mass loading of the wind
1.7~M$_{\odot}$~yr$^{-1}$ ({\em dashed line}; model U1-A),
17~M$_{\odot}$~yr$^{-1}$ ({\em solid lines}; U1),
49~M$_{\odot}$~yr$^{-1}$ ({\em dashed-dot-dot line}; U1-B), and
120~M$_{\odot}$~yr$^{-1}$ ({\em dotted line}; U1-C.).
\label{hist}
}
\end{figure}

Figure~\ref{hist}(a) shows the mass distribution of cool gas as a
function of velocity in X1-0, X1, and X1-2 at $t=0.27$ Myr. The
fraction of cool gas traveling at high velocities is low.  For
example, the mass travelling with  $v\ge500\mbox{ km s}^{-1}$ is less
than a few percent of the total shell mass, and the mass travelling
faster than the observed terminal velocity of 750~km~s$^{-1}$ is
$<0.1$\%. 

However, these results are best understood as upper limits. The
amounts of cool gas in the high-velocity end are progressively smaller
as the resolution increases.  Roughly factor of two decreases in the
mass of cool gas with velocities above 500~km~s$^{-1}$ occur between
runs with factor of two improvements in linear resolution.  This
suggests that we have not yet converged on the actual mass of
high-velocity gas, although we have set good upper limits.
We think this high-velocity gas is not likely to go away entirely,
even if we run a simulation with a higher resolution and with
additional physics. However, we can not fully quantify its amount with
our simulations.

The convergence properties of the peak velocity for cold gas
$v_{term}$ are somewhat better.  The highest resolution models show $<
20$\% variations, suggesting that the general result is reasonably
robust. The observed maximum value of 750~km~s$^{-1}$ is consistent
with the models at angles $\theta<60^{\circ}$.

\subsubsection{Mass Loading}

\begin{figure}
\includegraphics[width=\textwidth]{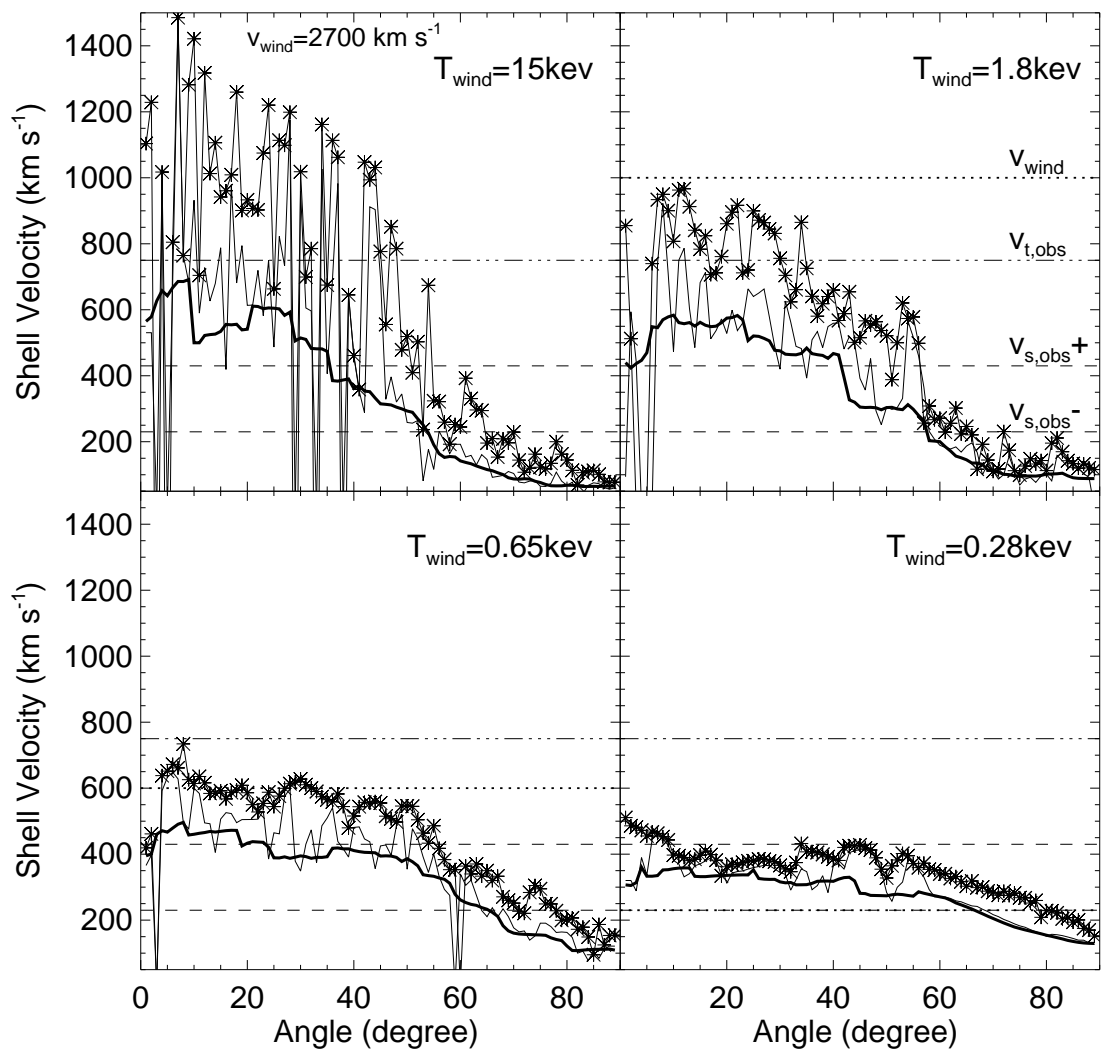}
\figcaption{Same as in Figure~\ref{term} for 
models with increasing mass-loading of the wind 
1.7~M$_{\odot}$~yr$^{-1}$ ({\em top left}; model U1-A) at $t=0.22$
Myr, 
 17~M$_{\odot}$~yr$^{-1}$ ({\em top right}; U1) at $t=0.27$ Myr,  
 49~M$_{\odot}$~yr$^{-1}$ ({\em bottom left}; U1-B) at $t=0.35$ Myr, 
and 
120~M$_{\odot}$~yr$^{-1}$ ({\em bottom right}; U1-C) at $t=0.41$ Myr.  
\label{termt1}
}
\end{figure}

Figure~\ref{termt1} shows the distributions of $v_{av}$, $v_{av,10}$,
and $v_{term}$ for simulations with different mass-loading rates and
so different wind terminal velocities: U1-A, U1, U1-B, and U1-C.  Cool
gas has higher terminal velocities in bubbles with hotter winds that
themselves have higher velocities $v_{wind}$.  The hottest wind, in
model U1-A, has $v_{wind}=2700\mbox{ km s}^{-1}$, while the coldest
wind, in U1-C, has $v_{wind}=250\mbox{ km s}^{-1}$.
Figure~\ref{hist}(b) shows the mass distribution of cool gas as a
function of velocity for the four runs at the time of blowout. The
cool gas with high $v_{term}$ carries only a small amount of mass (see
\S~\ref{subsub:term-res}).  For example, the fraction of cool gas with
$v>500\mbox{ km s}^{-1}$ is $\la 2\%$ of the total shell mass, and the
fraction with $v>750\mbox{ km s}^{-1}$ is $\la0.5\%$.  The bulk of
swept-up and cooled gas is driven to $v\sim400\mbox{ km s}^{-1}$ by
thermal pressure of hot sonic gas, but a small fraction of it seems to
be accelerated to higher velocity by the ram pressure of the same hot
gas as it accelerates to supersonic velocities during blowout.

\subsubsection{Mechanical Luminosity}

The terminal wind velocity (equation~\ref{ewind}) in our models X2 and
X3 with lower $L_{mech}$ is the same as that of our fiducial model X1, 
because the ejected mass $M_{SN}\propto L_{mech}$, and we keep the
mass-loading factor $\xi$ constant. In all three models
$v_{wind}\approx1000\mbox{ km s}^{-1}$.  

Figure~\ref{lwt} shows that the terminal velocities of cool gas are
not as high as the observed average terminal velocity,
$v_{t,obs}=750\mbox{ km s}^{-1}$ at most sightlines if
$L_{mech}\le10^{42}\mbox{ erg s}^{-1}$. It is much harder for the wind
ram pressure to accelerate the outermost shells and their fragments to
very high velocity, $>500\mbox{km s}^{-1}$ if their starting velocity
at blowout $v_b \la200\mbox{ km s}^{-1}$.

\subsubsection{Summary}

In summary, the terminal velocity of cool gas is determined by the
combination of total mechanical power $L_{mech}$ and wind terminal
velocity $v_{wind}$ (determined by the mass-loading rate $\xi$ in our
model). Bubble thermal pressure accelerates the bulk of shells to
their final velocity before blowout and ram pressure of the hot
transonic wind after blowout accelerates a small fraction of the cool
gas to much higher velocities.  Faster, less mass-loaded winds
accelerate cool gas to higher velocities.  Our wind is an
energy-driven wind, not a momentum-driven wind (Murray et al. 2005).
The thermal energy of the hot wind is gradually turning into its
kinetic energy by blowout (recall radiative cooling is turned off
inside the wind: see \S~\ref{sec:numerics}).  Our model accounts for
the observed $750\mbox{ km s}^{-1}$ terminal velocity of Na~{\sc i}
absorbing gas without invoking any additional physics such as
radiation pressure.

\subsection{Absorption Profiles}
\label{subsec:profiles}

To directly compare to observed profiles, we generate Na~{\sc i}
$\lambda 5890$ absorption line profiles along sightlines through our
simulations, as well as generalizing this procedure to fully model the
observed doublet Na~{\sc i} $\lambda,\lambda 5890,5896$. To generate the
profiles, we begin with the line intensity
\begin{equation}
I_{\nu}=I_{\nu}(0) \exp^{-\tau_{\nu}}=I_{\nu}(0) \prod_{i=1}^N \exp^{-\tau_i}
\label{I}
\end{equation}
where the optical depth through cell $i$ at frequency $\nu$ is
$\tau_{i}$, the sightline intersects $N$ cells, and the background
continuum is $I_{\nu}(0)$.  We normalize the continuum by setting
$I_{\nu}(0)=1$ and compute the profile as a function of the
macroscopic velocity of Na~{\sc i} absorbing gas, $v$. We compute
$I_v$ in the simulations by setting $v=c(\nu-\nu_0)/\nu_0$, and
computing the optical depth contributed by each cell in each of 1000
velocity bins.  The optical depth in each cell
\begin{equation}
\tau_{i}(v)=N_{NaI}~s~\lambda_{NaI} P(\Delta v), 
\label{tau}
\end{equation}
(Spitzer 1978) where $N_{\mbox{Na~{\sc i}},i}$ is the column density
of Na~{\sc i} in a cell $i$, and the absorption cross section
$s_{\nu}$ integrated over frequency $\nu$ for $h\nu\gg kT$ is $s
\approx 2.654\times10^{-2}f_{5890}$ with the oscillator strength
$f_{5890}=0.6$.  The Maxwellian velocity distribution function,
$P(\Delta v)$, is
\begin{equation}
P(\Delta v)=\frac{1}{\sqrt{\pi}b}\exp{-(\Delta v^2/b^2)}, 
\end{equation}
with $\Delta v=v-v_i$ and $b=\sqrt{2k_B T_i/\mu m_H}$ for thermal
broadening.  Our simulations do not track chemical abundances,
ionization state, or dust depletion, so we do not directly predict the
Na~{\sc i} column.  For the purpose of illustration, we compute the
total Na~{\sc i} column from the total H~{\sc i} gas column using the
conversion used by Martin (2005) to estimate $N_{\mbox{H~{\sc i}}}$
from their $N_{\mbox{Na~{\sc i}}}$ measurements, $N_{\mbox{Na~{\sc
i}}} = 1.122\times10^{-6}N_{\mbox{H~{\sc i}}}$.

\begin{figure}
\includegraphics[width=\textwidth]{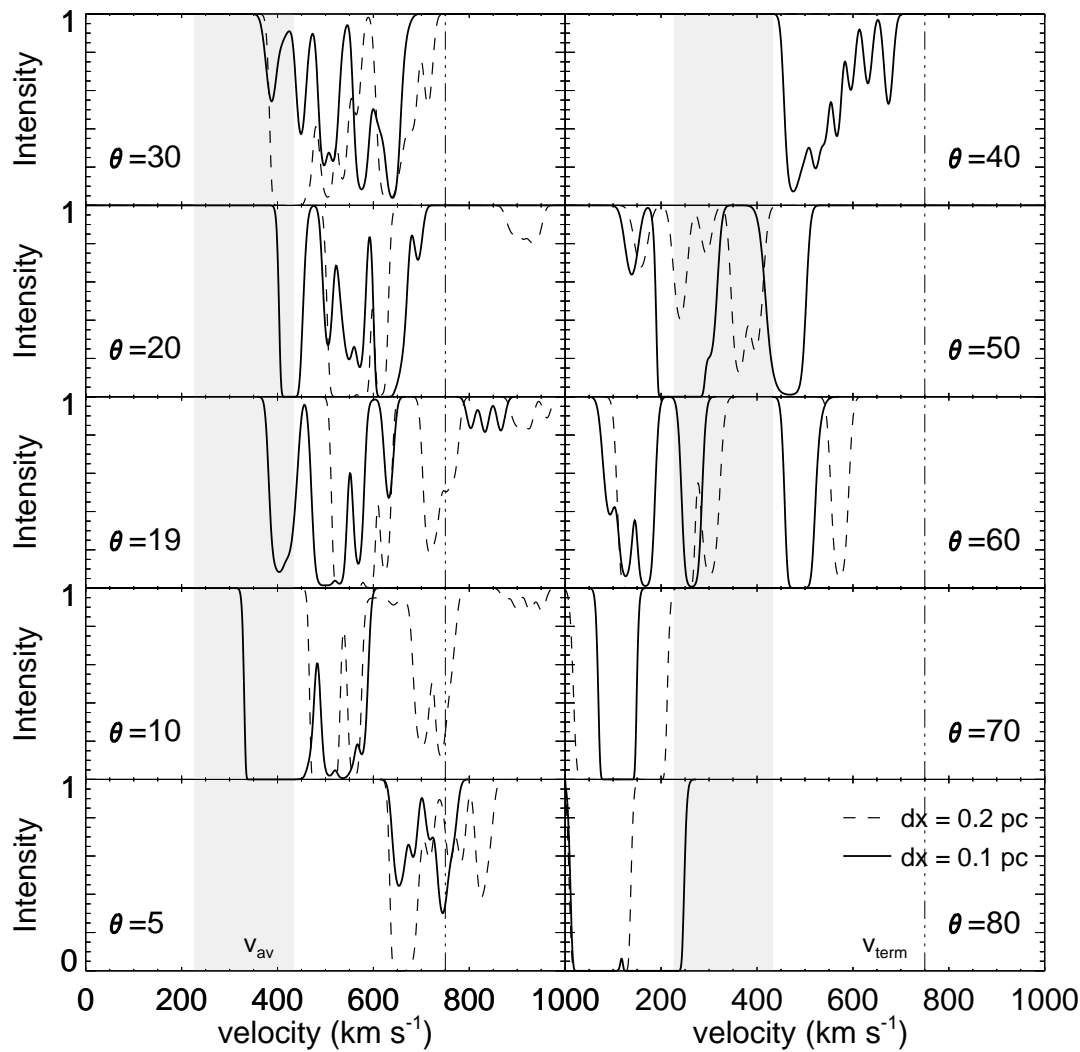} 
\figcaption{Simulated Na~{\sc i} 5890 absorption profiles at
sightlines $\theta=5^{\circ}$, $10^{\circ}$, $19^{\circ}$,
$20^{\circ}$, $30^{\circ}$, $40^{\circ}$, $50^{\circ}$, $60^{\circ}$,
$70^{\circ}$, and $80^{\circ}$ ({\em clockwise from the bottom left})
for models with $dx = 0.1$~pc ({\em solid line}; model X1-0) and 0.2
pc ({\em dashed line}; X1).  We also show the observed range of
average shell velocity in {\em shaded area} and the observed average
terminal velocity in {\em dash-dot-dot} line.
\label{NaI}
}
\end{figure}

We generated Na~{\sc i} 5890 line profiles along lines of sight
through models X1 and X1-0 to show the effect of numerical resolution
on the line profiles in Figure~\ref{NaI}.  Each sightline is described
by its inclination from the polar axis of the simulation.  The profile
shapes reflect the different structure in these simulations.  We can
for example compare the absorption profiles in the middle left panel
in Figure~\ref{NaI} with the density and velocity distributions of
shells in Figures~\ref{line} and \ref{bline}. The sightlines at
$\theta=19^{\circ}$ intersect three and four shell fragments in X1 and
X1-0 respectively. Each of these fragments generates a $\Delta v \ga
50-100~\mbox{km s}^{-1}$ absorption line.  A few of these lines are
optically thick with the line profile becoming completely black at the
line center. As the wind expands, column densities will drop linearly
with radius due to geometric dilution.  The complex of lines is spread
out over $\sim400~\mbox{km s}^{-1}$.

Figure~\ref{NaI} shows that the line profiles from the high and low
resolution models present very similar structure for the most part.
One minor exception is the minimum velocity.  From $10\deg$ to $20\deg$, the
line profiles show a sharp cut-off at a velocity $\sim500\mbox{ km
s}^{-1}$ in X1, but $\sim3-40\mbox{ km s}^{-1}$ in X1-0. 
This cut-off reflects the
velocity of the first R-T fragments that form from the accelerating
swept-up shells, which is lower in the high-resolution simulation.
The overall lineshape does appear reasonably well converged.

Observers see the average over parallel sightlines subtending a few
kiloparsecs of the disk.  They must also contend with the instrumental
response function, and the doublet nature of the Na~{\sc i} line,
which has components at 5890 \AA\ and 5896 \AA, with optical depths
differing by a factor of $\sim 2$. To compare to the actual observed
profiles, we generate the Na~{\sc i} 5890/5896 doublet in the frame of
Na~{\sc i} 5890 by assuming the ratio of equivalent widths is only
1.3, typical of the observed lines in ULIRGs (rather than the
optically thin limit of two).

We present average line profiles of the doublet over a $20\deg$
wide set of radial sightlines spaced at degree intervals.  Although
this is not exactly what is measured, it is comparable because these
rays will subtend about 1-2~kpc of the shell when the bubble is $\sim
10\mbox{ Myr old}$, so they intersect the same region of the shell,
albeit at slightly different angles.  We are actually doing the
analysis at an earlier time, when the bubble is a factor of 100
smaller, and correspondingly higher column density, though, so we also
must apply a geometric dilution factor to the column densities.

\begin{figure} 
\includegraphics[width=\textwidth]{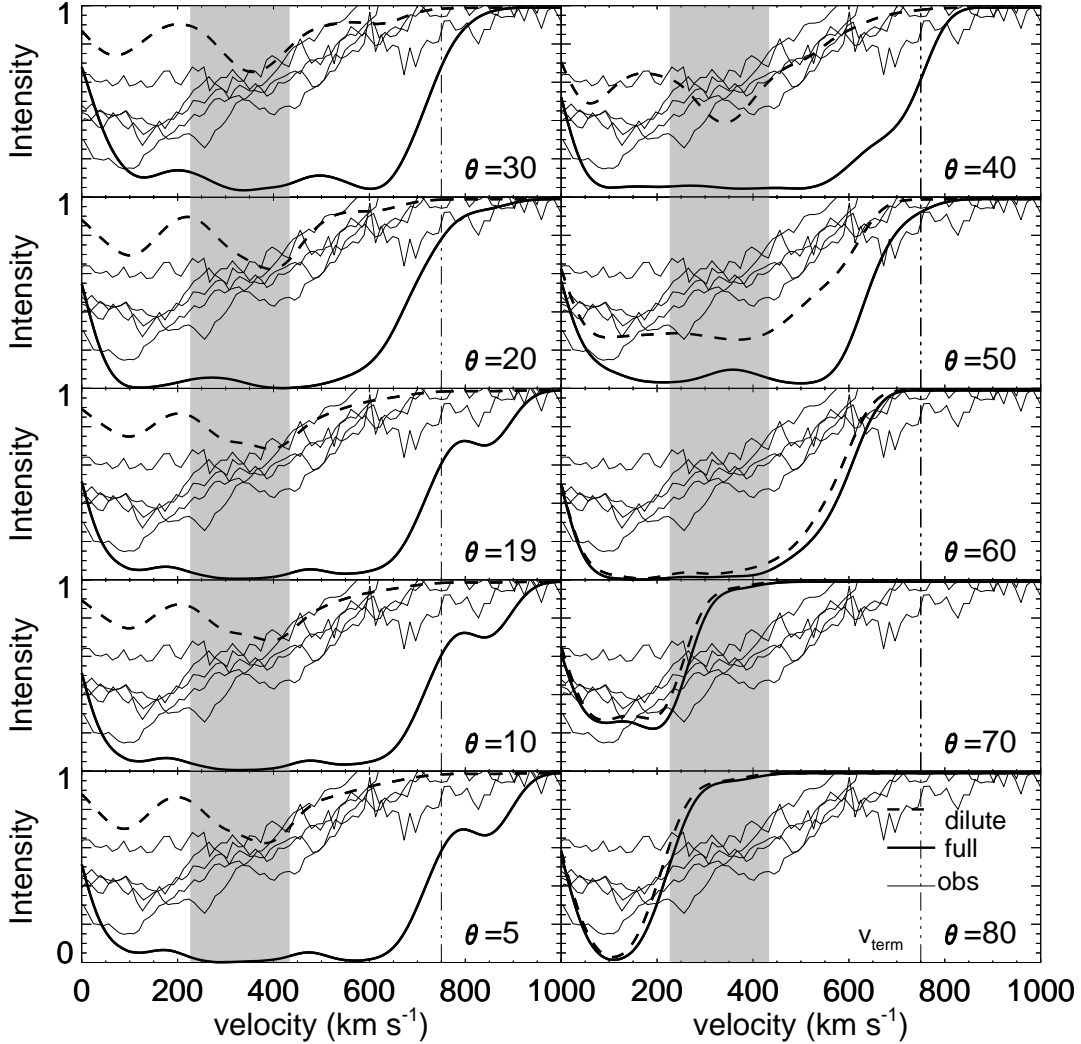} 
\figcaption{Simulated Na~{\sc i} 5890/5896 doublet absorption profiles
  averaged over $ª£20^{\circ}$ centered on the given (non-uniformly
  distributed) angles for model X1-0 with $dx = 0.1$~pc ({\em solid
    line}). The same profiles are shown after geometric dilution of
  the column densities by a factor of 100 ({\em dashed line}), and
  application of a Gaussian instrumental broadening with
  FWHM$=65$~km~s$^{-1}$, for comparison with five observed ULIRG
  spectra ({\em thin solid lines}) from Martin (2005).  Note that the
  velocity frame is centered on the 5890 line; blue-shifted absorption
  from the 5896 line lies at low velocities in this frame.  We also
  show the observed range of average shell velocity in {\em shaded
    area} and the observed average terminal velocity in {\em
    dash-dot-dot} line.
\label{NaI20} 
} 
\end{figure}

Figure~\ref{NaI20} compares our simulated doublet line profiles, with
and without geometric dilution of the column densities by a factor of
100, to five typical ULIRG spectra from Martin (2005) that have an
instrumental broadening with FWHM$\simeq 65$~km~$^{-1}$. We convolve
in a Gaussian of that width to simulate the broadening.  The widths of
the absorption profiles are very similar between our model at
$\theta\le50^{\circ}$ where blowout occurs and the observations.  The
undiluted absorption in our model much exceeds that observed.  Our
model overpredicts the cool gas column for two reasons (see
\S\ref{subsec:coolgas}). First, because we analyzed the models at a
very early time in the starburst.  At later times, the spherical
expansion of the wind and the cold gas embedded within it dilutes
their column densities. Second, the fragmentation of the shell is
incompletely resolved, so some material remains cool that in reality
would have been mixed into the hot wind. We account for the first
factor by directly reducing the column densities, producing the
diluted profiles shown in the Figure, which reproduce the observed
intensities well.

The ULIRG spectra show absorption beginning from the systemic
velocity, and extending to high velocities. Although this zero
velocity absorption is absent in Figure~\ref{NaI}, it is seen in the
simulated doublet profiles shown in Figure~\ref{NaI20}, as it comes
primarily from contributions from the $\lambda$5896 line of the
doublet blueshifted into zero velocity of the $\lambda$5890 line.
(Note our model does not include the contribution from stellar
absorption).

\subsubsection{Long Slit Observations}

Measured Doppler shifts across ULIRGs show the cool outflow is extended
spatially;  and some outflows present a significant velocity gradient
over kpc scales (Martin 2006). 
Our simulation ends long before the wind has reached such large scales
and does not include the rotation of the galactic disk. The simulation 
does demonstrate, however, the amplitude of the velocity gradient
that arises across the minor axis due to projection effects.

\begin{figure}
\includegraphics[width=\textwidth]{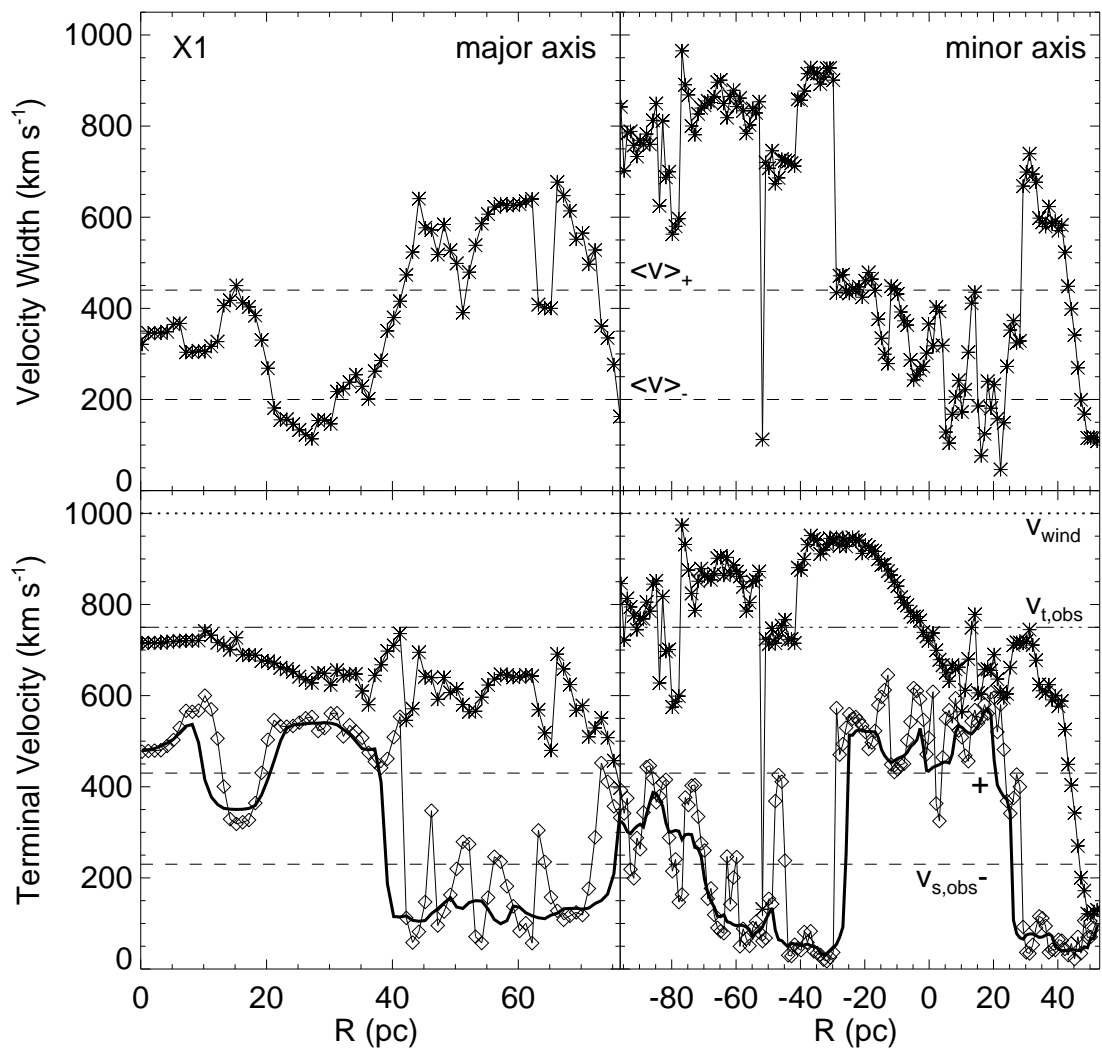}
\figcaption{The velocity width is plotted as in Figure~\ref{lw} ({\em
    top panels}), and the mass-weighted average velocity and the
    terminal velocity are plotted as in Figure~\ref{term} ({\em bottom
    panels}) for model X1.  Parallel sightlines are chosen along a
    slit oriented at $\theta=30^{\circ}$ from the axisymmetric axis
    and the major axis ({\em left panels}) and the minor axis ({\em
    right panels}).
\label{30}
}
\end{figure}

The wind is launched without any net angular momentum in our 2D simulation.  
We examined whether, in the absence of rotation, the
position affected the velocity much. 
  Figure~\ref{30} shows terminal and mass-weighted average velocity
  along slits oriented at $30^{\circ}$ from the major and minor axes
  of the disk, using parallel lines of sight through a 3D rotation of
  the 2D simulation. 
The absorption properties along the major axis of the galaxy will be
symmetric about its center.  An asymmetric velocity gradient is
produced along the minor axis because one side of the outflow cone is
directed more along the sightline.  Substantial variations in velocity
width and average velocity are seen at $\sim40$ pc and $\pm30$
pc. These occur where sightlines intersect massive, slow-moving shells
at $\theta>45^{\circ}$ (see Figure~\ref{term}) as well as light,
fast-moving blowout fragments at $\theta<45^{\circ}$. On the other
hand, very high terminal velocities are seen on most sightlines since
they intersect the fast blowout components.

The simulated  sightlines encounter both fast-moving and slow-moving gas at
various latitudes, unlike the radial sightlines that we studied in previous
sections. Our simulations viewed along non-radial sightlines 
naturally account for a wide velocity range of cool gas starting at
$v\approx0$ and a high terminal velocity.

\subsection{Line Widths of Na~{\sc i} Absorbing and H$\alpha$ Emitting Gas}

Finally, we crudely try to separate low-ionization, Na~{\sc i} absorbing gas 
and ionized H$\alpha$ emitting gas in the cooled shell fragments in our simulations. 
Our aim is to compare the line widths of both components.
Heckman et al.\ (2000) observed no correlation between the absorption
and emission line widths for these lines, and drew the conclusion that
only one of the lines could originate in the swept-up shells. 

To model the H$\alpha$ emission, we must decide whether the emitting
gas is primarily photoionized by the central starburst or shock
ionized in the wind.  Shock ionization dominates at least some
starburst-driven winds, such as NGC~1482 (Veilleux \& Rupke 2002).
Some observed ULIRGs also show extended shock-excited nebulae
(Monreal-Ibero et al.\ 2006). However, the dynamics observed by
Heckman et al.\ (2000) seem unlikely to come from shock ionized gas,
since they see no correlation with the motions of the cold gas that
presumably trace the shock.  Therefore we assume photoionization by
the central starburst, choosing an ionizing photon luminosity $Q$ and
a density distribution from a time the galaxy is likely to be
observed.  Figure~\ref{lmech} shows the ionizing photon luminosities
as a function of time for three starburst cases that we consider.
Since the propagation of ionization fronts critically depends on both
the densities and the positions of shells, we can not extrapolate the
density distribution at blowout to $t\approx10$ Myr using the
ballistic approximation. We can, however, still vary the strength of
photon luminosity over a few orders of magnitude to examine the effect
of attenuating the photon flux by $1/r^2$ on our existing simulations.
For example, a shell will experience about three orders of magnitude
less photons per unit area after it travels with $500\mbox{ km
  s}^{-1}$ for 10 Myr from $r_b=200$ pc. For our purpose of
demonstrating the lack of correlation between the line width of
Na~{\sc i} absorbing and $H\alpha$ emitting gas, this crude method is
sufficient.

To solve for the transfer of ionizing radiation across our grids, we
use the photon-conserving radiative transfer code developed by Abel et al.\ (1999) in the same manner as it was used in Fujita et al.\ 
(2003).  It computes the propagation of ionization fronts around a point
source; in our study this is a central starburst source. This code is
a post-processing step that operates on a given density distribution
at a given time step in our simulation. However, the ionization fronts
propagate sufficiently rapidly to adjust almost instantaneously to a
changing density distribution.

\begin{figure} 
\includegraphics[width=\textwidth]{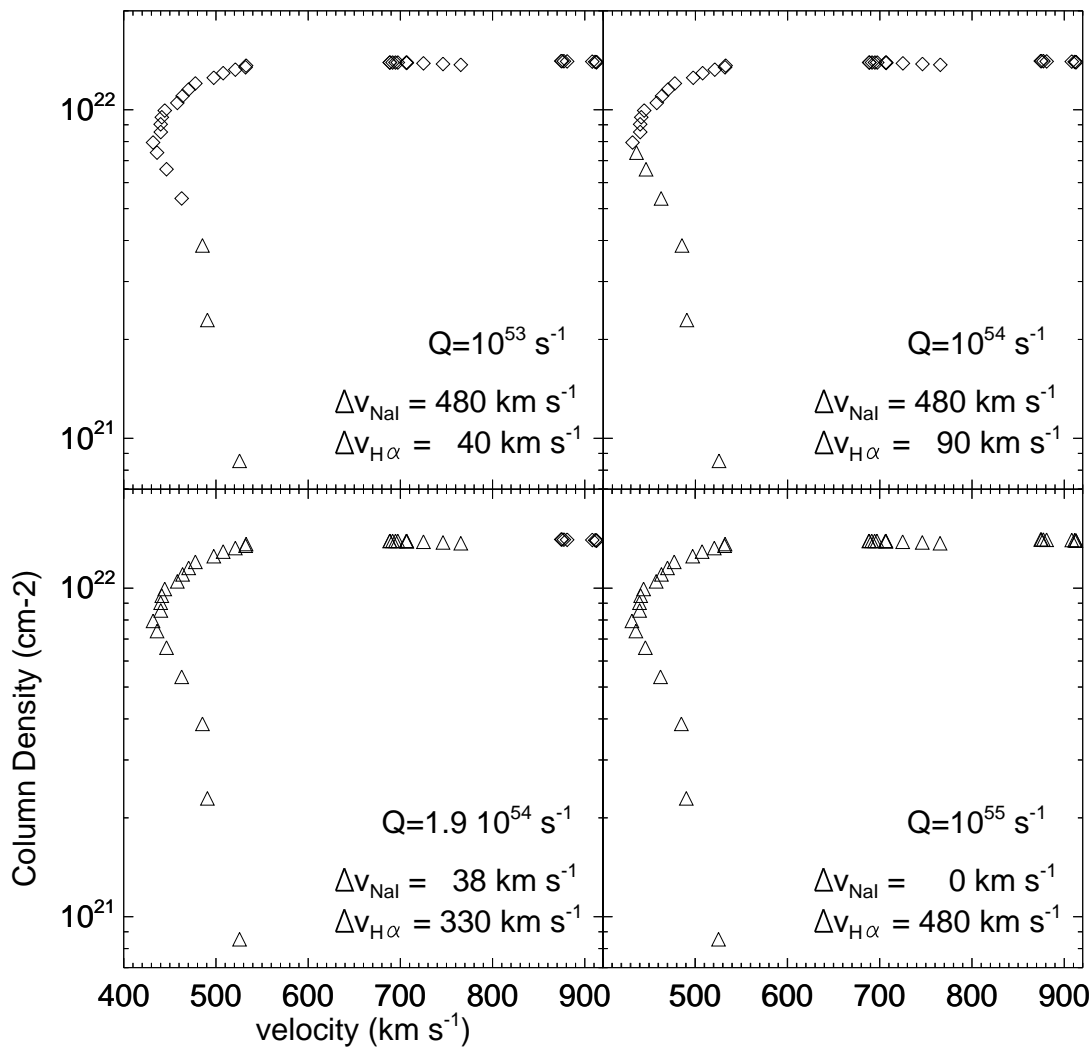} 
\figcaption{The column density distributions as a function of radial
velocity for Na~{\sc i} absorbing gas ({\em diamonds}) and H$\alpha$
emitting gas ({\em triangles}) with photon luminosities of
$Q=10^{53}$, $10^{54}$, $1.9\times10^{54}$, and $10^{55}$ photons
s$^{-1}$ in a line of sight through the center at an angle of
13$^{\circ}$ from the vertical axis in model U1.
The velocity widths of the two components vary as Q is changed.
\label{ion} } \end{figure}

Figure~\ref{ion} shows the column density as a function of velocity
for Na~{\sc i} absorbing gas ({\em diamonds}) and H$\alpha$ emitting
gas ({\em triangles}) in run U1 at $\theta=13^{\circ}$ with ionizing
luminosities of $Q=10^{53}, 10^{54}, 1.9\times10^{54}$, and $10^{55}$
photons s$^{-1}$.  With $Q\le10^{54}$~photons~s$^{-1}$, the line
widths are very small for H$\alpha$ emitting gas, $\Delta v_{H\alpha}<
100\mbox{ km s}^{-1}$, but large for Na~{\sc i} absorbing gas, $\Delta
v_{NaI}=480\mbox{ km s}^{-1}$.  The densest shell at $r=110$ pc traps
the photons rather effectively.  As Q is increased, that first shell
is ionized, but the second shell then traps the photons.  The {\em
  bottom left} panel shows that the Na~{\sc i} line width is now very
small, $\Delta v_{NaI}=38\mbox{ km s}^{-1}$, while $\Delta
v_{H\alpha}=330\mbox{ km s}^{-1}$.  With $Q\ga 2\times10^{54}$, all
the shells are ionized.  In reality, the transition from large $\Delta
v_{NaI}$ to $\Delta v_{NaI}\approx0$ should not be this abrupt because
many combinations of fragments and clumps are possible within the
observers' $10^{\circ}\times10^{\circ}$ field of view.  As we note, we
are far from presenting realistic distributions of Na~{\sc i}
absorbing and H$\alpha$ emitting gas. However, Figure~\ref{ion} still
demonstrates that the line widths of Na~{\sc i} absorbing and
H$\alpha$ emitting gas can easily be uncorrelated although they both
originate in the swept-up shells.

\section{Caveats and Summary}

\subsection{Caveats}

This study clearly is not the final word on this subject.  Rather it
is a proof of the concept that a starburst wind can accelerate neutral
gas up to high velocities without special circumstances, and that the
resulting flow configuration can reproduce many of the observable
properties of ULIRG winds.  We here recap the approximations we have
made in order to treat this problem, and add some discussion of how
they might be lifted.

The biggest approximation of our study is that we analyze the
kinematics of shell fragments at $t\ll 1$ Myr in our small grids and
extrapolate the results by the ballistic approximation including
geometric dilution for comparison with the observations at $t \gg 1$
Myr.  Much of the total mechanical energy from even an instantaneous
burst of star formation has yet to be deposited at an age of 2~Myr
when our simulation ends.  These dense clumps may further fragment and
ultimately be destroyed by R-T and Kelvin-Helmholtz instability as the
high-velocity flow of gas from within the bubble streams through them.
Conversely, they may be dense enough to be accelerated further prior
to their destruction by the ongoing starburst wind.

As the grid resolution of our models improves to 0.1 pc, we do begin
to resolve the details of hydrodynamic instabilities.  Marcolini et
al.\ (2005) used 0.1~pc resolution for their simulations of the
dynamical shredding of radiatively cooling clouds, and seem to have
reached adequate resolution.  Thus, in future work, continuing to
evolve our bubbles in bigger grids with the same resolution may well
improve our results.

We have made four other substantial physical approximations as
well. First, the assumption of aziumthal symmetry further suppresses
R-T instability. Mac Low et al.\ (1989) pointed out that the typical
spike and bubble structure of the R-T instability is limited to rings
in an axisymmetric blowout.  The detailed distribution of the
fragments will certainly be different in three dimensions, but finer
fragments will probably have a broader velocity range, and thus are
unlikely to change our qualitative result.  The behavior of individual
fragments will also not be qualitatively changed by the
dimensionality, although two-dimensional fragments split into larger
pieces initially (Stone \& Norman 1992b; Korycansky et al.\ 2002).

Second, we neglected magnetic fields in this study.  We showed that
magnetic fields reduce the peak shell density, and thus the
fragmentation (\S~\ref{subsec:effects-res}), but that they don't act
to do so before fragmentation is essentially complete in our model, so
that this effect is secondary.  Magnetic fields can, on the other
hand, help preserve individual fragments from further breakup (Stone
\& Norman 1992b; Mac Low et al.\ 1994; Shin et al.\ 2008), possibly
even allowing their further acceleration in the continuing starburst
wind.
  
Third, thermal conduction is not explicitly treated in our model.
This acts on parsec scales, so numerical diffusion and turbulent
mixing will still dominate over thermal conduction under most
circumstances at the resolutions that we consider.  It is worth
noting, though, that studies of individual clump fragmentation have
found that thermal conduction can have stabilizing effects (Marcolini
et al.\ 2005; Vieser \& Hensler 2008).  Fourth, the dynamical effects
of photoionization have also been neglected.  This could heat at least
low column-density shells and fragments up to $10^4$~K, reducing their
density contrast with the wind and enhancing their tendency to
fragment.

We also assume a single starburst at the disk center for modeling a
galactic outflow. In realistic galaxies, we expect multiple star
clusters to be scattered around the disks (e.g.\ Vacca 1996; Martin
1998). A more complicated structure of cool shells and their fragments
is expected with a realistic distribution of star formation as shown
by models of supernova-driven turbulence in our own galaxy (e.g
Avillez 2000; Avillez \& Berry 2001; Joung \& Mac Low 2006). Such a
distribution was modeled with a fractal density distribution by Cooper
et al.\ (2008).  Fragile et al.\ (2004) showed that in dwarf galaxies,
distributed supernova explosions resulted in increased transfer of
energy to the cold gas compared to the centrally concentrated energy
injection assumed here and by Mac Low \& Ferrara (1999), so our
approximation likely gives a lower limit to the kinetic energy of the
cold gas. This supports our result of a wide velocity range in Na~{\sc
  i} absorbing gas.  The next obvious step is to study the effects of
realistic star formation on the shell kinematics in three dimensional,
AMR simulations.

Another related issue is that our models assume a low-density,
uniform, gas distribution above the site of blowout. However, ULIRGs
are formed in galaxy mergers. The actual situation near the nucleus of
merging galaxies is likely to be more chaotic.  We rely on the idea
that the large-scale tidal tails and other merger structures are
generally going to be well removed from the wind generation region and
will not cover a large fraction of the sky.  Ultimately, to remove
this limitation, full models of merging galaxies at high resolution
will be needed, but this remains some years in the future.

\subsection{Summary}
    
We study the origin and the kinematics of cool gas that produces
Na~{\sc i} absorption lines in galactic winds, using hydrodynamic
simulations of the blowout of starburst-driven superbubbles from the
molecular disk of ULIRGs.  The bubbles sweep up the dense disk gas,
which quickly cools to form dense, thin shells.  The cooled shells
fragment by R-T instability as they accelerate in the stratified
atmosphere of the disk.  This blowout occurs very early in our models,
at $t \ll 1$ Myr in models with $L_{mech}\ge10^{41}\mbox{ erg
s}^{-1}$.  The dense fragments left by the R-T instability lag behind
the low-density, high-velocity wind.  These fragments carry most of
the mass that is swept up by the bubbles, and should have the highest
column of Na~{\sc i}.  The results of our numerical convergence study
suggest that superbubble blowout, combined with subsequent geometrical
dilution, can reproduce not just qualitative but quantitative
properties of the observed lines.

As a result of R-T fragmentation, multiple shell fragments and clumps
travel at different velocities. A sightline going through them
reproduces the observed broad line width of the Na~{\sc i} absorption
profiles: $\langle v \rangle=320\pm120\mbox{ km s}^{-1}$.  This result does not
appear to depend strongly on physical parameters such as mass-loading
rate, mechanical luminosity, or surface density.  However, this result
requires sufficient numerical resolution to follow secondary
fragmentation of the shell, so that any line of sight runs over
multiple cold gas fragments.  We find that a resolution of at least
$\sim0.2$ pc is required to produce this effect, if mass-loading rates
remain moderate $\xi\la10$. The suggestion by Heckman et al.\ (2000) that 
swept-up shells will not show a wide range in velocity is based on a hydrodynamic
simulation of M82 with a resolution of 4.9 pc (Strickland \& Stevens
2000), more than an order of magnitude worse. 

The mass-weighted average velocity of cool gas in galactic outflows is
found to be $\sim400$--500~km~s$^{-1}$ among all the models with
$L_{mech}=10^{43}\mbox{ erg s}^{-1}$, but $\la200\mbox{ km s}^{-1}$ in
models with $L_{mech}=10^{42}$ and $10^{41}\mbox{ erg s}^{-1}$.  
The bulk of shells and their fragments are accelerated by the thermal
pressure of the bubble interior gas, which is proportional to
$L_{mech}$. No other parameters influence the results.  The
mass-weighted average velocities in our high-resolution simulation
with $L_{mech}=10^{42-43}$~erg~s$^{-1}$ agree with the observed
value $v_{s,obs}=330\pm100\mbox{ km s}^{-1}$.  A mechanical luminosity
of $L_{mech}=10^{43}\mbox{ erg s}^{-1}$ corresponds to an
instantaneous burst of $M_*=10^9~\mbox{ M}_{\odot}$ or a continuous
SFR$=500~\mbox{ M}_{\odot}\mbox{ yr}^{-1}$ at $t<1$ Myr after the
onset of starburst.

As the swept-up shells fragment by R-T instability, the bubble
interior gas blows out and becomes a low-density supersonic wind, as
its thermal energy is converted to kinetic energy by expansion.  The
ram pressure of this wind continues to accelerate entrained cold
fragments, and to sweep up high-altitude disk gas, producing small
amounts of cold gas with velocities of the order of the terminal
velocity of the wind.  The correlation between wind terminal
velocities and terminal velocities of cool gas in simulations with
different mass-loading rates supports this picture.  If the wind
velocity reaches $v_{wind}\ga1000\mbox{ km s}^{-1}$, a significant
fraction of sightlines through a simulation encounters a terminal
velocity close to or greater than the observed average terminal
velocity of Na~{\sc i} absorption profiles $v_{t,obs}=750\mbox{ km
  s}^{-1}$.  The mass of cool gas with such high velocity is less than
a few percent of that of the total cool gas, however.  These outermost
shells and fragments with high velocity do fragment further in our
highest-resolution simulation, but their fragments are traveling with
similarly high velocity.  Although the mass in the high-velocity tail
decreased as the resolution is increased, we believe it is not likely
to go away entirely in a higher-resolution simulation.

The clumps and fragments seen in the simulations may be observed as
Na~{\sc i} absorbing gas or H$\alpha$ emitting gas, depending on the
amount of ionizing photons produced from the central starburst source.
To study this, we used a ray-tracing method to model the location of
the ionization front in our models as a function of time, and so to
trace the gas emitting in H$\alpha$.  By varying the photon
luminosity, we showed that the velocity range of the ionized and
neutral components do not show any correlation with each other. Thus
the lack of correlation in the observed line widths of Na~{\sc i}
abosorbing gas and H$\alpha$ emitting gas does not rule out the
swept-up shells as the origin of both components.

The hot wind in our model is purely energy-driven. By construction the
interior bubble can never become momentum-driven because radiative
cooling is turned off in the bubble interior in our models.   Future work must
determine whether this mechanism reproduces the empirical correlation
observed between the maximum outflow velocity and the escape velocity
of the host galaxy (Martin 2005). It should also predict the scaling of the
mass-loss efficiency with galaxy mass, a key input in cosmological
models (e.g.\ Oppenheimer \& Dav\'e 2006, 2008)

\acknowledgements We thank M. L. Norman and the Laboratory for
Computational Astrophysics for use of ZEUS-3D, R. F. Coker, G. Gisler,
R. M. Hueckstaedt, and C. Scovel for the help with getting started
with SAGE, and M. K. R. Joung for useful discussions. We thank the
referee, D.\ Rupke, for his extensive assistance in placing this work
in context. Computations were performed on the SGI Origin 2000 of the
Rose Center for Earth and Space, and the QSC machine of Los Alamos
National Laboratory (LANL). AF was supported by a cooperative
agreement between UCSB and LANL.  CLM thanks the Alfred P. Sloan
Foundation and the David and Lucile Packard Foundation for support for
this work.  M-MML was partly supported by NSF grants AST99-85392 and
AST03-07854, and by stipends from the Max-Planck-Gesellschaft and the
DAAD.

\clearpage

\appendix

\section{Appendix: Blowout in a Dwarf Galaxy with ZEUS and SAGE}

A major issue in our simulations is whether we sufficiently resolve
the unstable shell during blowout.  To further examine this question,
in this Appendix we describe a comparison between models of blowout
from a dwarf galaxy similar to that described by Fujita et al.\ (2003)
run with ZEUS-3D, and single-grid and AMR versions of SAGE.

SAGE is an AMR hydrodynamic code developed at LANL/SAIC.  It
is second order accurate using a piecewise linear Godunov scheme
(Kerbyson et al.\ 2001; Gittings et al.\ 2008).

The dwarf galaxy has a nominal redshift $z=8$, and disk mass $M_d=10^8
\mbox{ M}_{\odot}$.  We choose the midplane density of this galactic
disk so that it has an exponential surface density profile.  We set up
the gas in hydrostatic equilibrium with a background dark matter halo
potential, with an effective turbulent velocity of $10\mbox{ km
  s}^{-1}$.  Mass and energy are injected at the center of the disk,
corresponding to a mechanical energy of starburst, $10^{40} \mbox{ erg
  s}^{-1}$, with a mass-loading rate of 0.1 $\mbox{ M}_{\odot} \mbox{
  yr}^{-1}$.  See the details of disk and halo parameters in Fujita et
al.\ (2003).  

In SAGE, the cooling is solved with an explicit method that subcycles
the cooling source terms, based on a radiative cooling routine that
solves for non-equilibrium chemistry (Abel et al.\ 1997).  The
timesteps for each subcycle are determined as $\epsilon e/\dot{e}$
where $e$ is internal energy density and $\epsilon=0.1$.  In ZEUS-3D,
on the other hand, we assume that the cooling rate is a function of
temperature only, using calculations of equilibrium ionization cooling
rates by Sutherland \& Dopita (1993) with a cut-off at $T < 10^4$ K,
and ignore inverse Compton cooling for simplicity.  The cooling curve
is implemented in the energy equation with a semi-implicit method,
using a Newton-Raphson root finder.

In order to avoid overcooling in the bubble interior (see \S 4), we advect two materials
independently in SAGE; 1) disk and halo gas and 2) hot, metal-enriched
gas injected at the starburst site, 
and turn off cooling in the latter. SAGE computes hydrodynamics of a
multi-material flow, 
assuming all materials in a cell have the same pressure. 
The time required to process mixed cells is linearly dependent on the number of materials.
In ZEUS-3D we use the tracer field (Yabe \& Xiao 1993) to suppress
interior cooling.

We ran simulations with and without radiative cooling,
and with and without AMR.
Table~\ref{sz} shows the parameters for our computations. 
With SAGE we use both uniform and AMR grids 
having a maximum effective resolution of 0.0693~pc.
With ZEUS, we use a ratioed grid 
having the same central resolution as the lowest resolution SAGE runs,
and decreasing resolution outside the innermost $0.1\times0.2$
kpc$^2$. 

SAGE was run on the QSC cluster at LANL using 12 Alpha processors for
the UN, UC, AN, and AC runs, 24 processors for the BC run, and 48
processors for the CC run, while ZEUS-3D was run on a SGI Origin 2000
using 8 processors.  SAGE's performance is linear up to at least a few
hundred processors (Kerbyson et al.\ 2001), while the loop-level
parallelized performance of ZEUS-3D levels off after 8 processors (a
massively parallel, domain-decomposed version, ZEUS-MP, has linear
performance to over 512 processors).  The number of active cells used
for the AMR simulations with 5 and 6 levels of refinement is less than
the number of cells used for the ZEUS-3D simulations. The CPU time per
cycle is much smaller with SAGE compared to that with ZEUS-3D, if
cooling is included. ZEUS-3D spends the most time searching for
convergence around the cut-off temperature of the cooling curve, since
radiative cooling in our dense, high-redshift disk is very efficient.

\begin{figure}
\includegraphics[width=\textwidth]{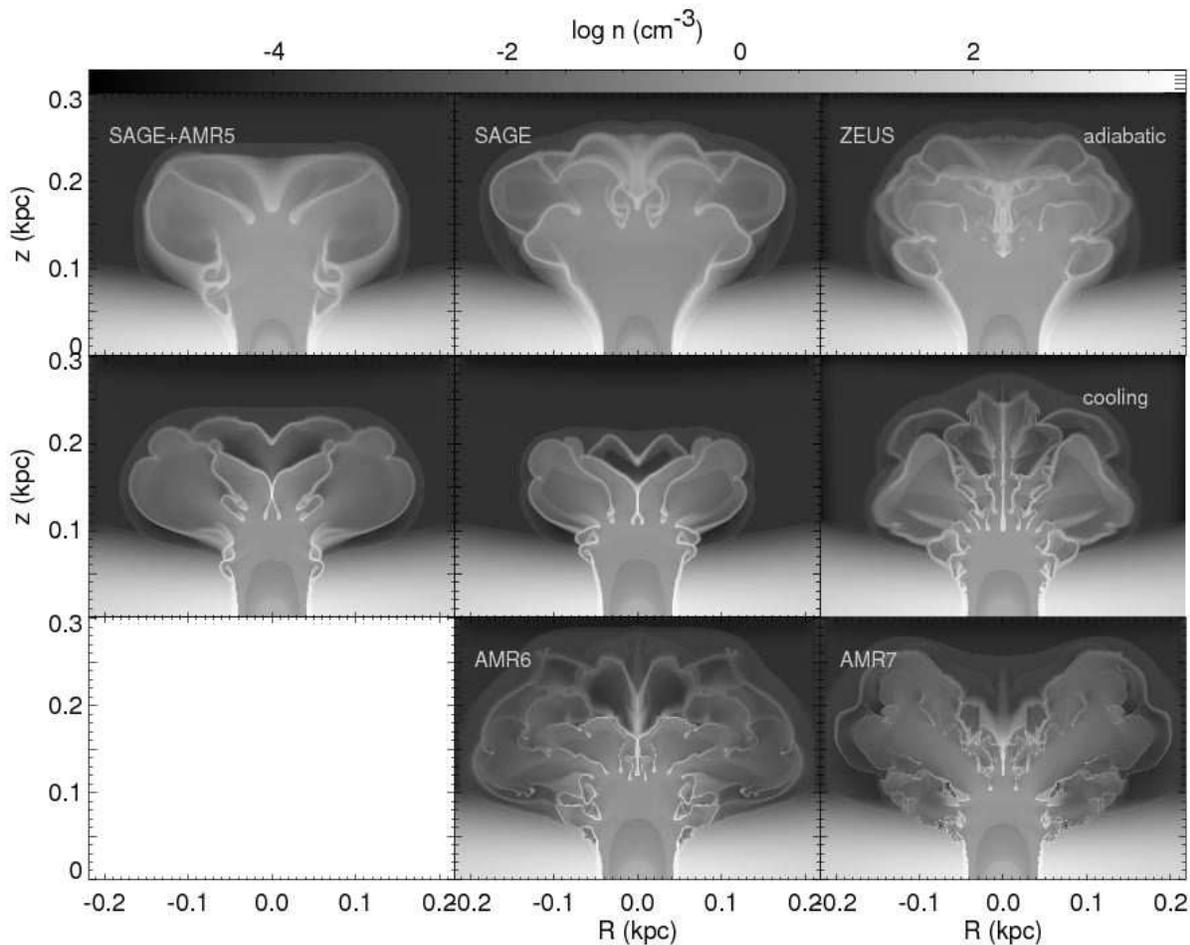} \figcaption{ The density
distributions at blowout in our model dwarf galaxy with $M_d=10^8
\mbox{ M}_{\odot}$ and $L_{mech}=10^{40}\mbox{ erg s}^{-1}$ at $z=8$.
The bubbles are shown at the time of blowout.  This is $t=1$ Myr for
models without radiative cooling, $t=1.3$ Myr for SAGE models with
radiative cooling, and $t=1.1$ Myr for ZEUS-3D models.  The top row
shows models without radiative cooling using SAGE with uniform grid
and 5 levels of AMR, and using ZEUS-3D.  These are models AN, UN, and
RN.  The middle row shows the same grids with radiative cooling,
models AC, UC, and RC.  Finally, the bottom row shows SAGE models with
cooling and 6 and 7 levels of AMR, models BC, and CC.
\label{compare}
}
\end{figure}

Figure~\ref{compare} shows the density distributions of our models at
the times when the bubbles blow out of the disks.  This time was
chosen so that the positions of outer shock fronts in the horizontal
direction agree.  Figure~\ref{compare} shows qualitative agreement
between the results with SAGE and ZEUS.  In our no-cooling runs, the
positions of outer shock fronts agree within $\sim2\%$.  Those of
inner shock fronts agree within $\sim5\%$ if SAGE is run in the
uniform grid and $\sim10\%$ if SAGE is run with AMR, because we chose
an AMR refinement criterion that did not act in that region, while the
uniform grid had the resolution of the highest AMR refinement level.

The linear piecewise advection method of SAGE seems more diffusive
than the Van Leer (1977) method of ZEUS.  However, we show in
Figure~\ref{compare} that the SAGE/AMR run begins resolving fine
structures caused by R-T instability as well as ZEUS-3D run when one
more level of refinement is added, doubling the maximum effective
resolution.  Even in this case, the SAGE/AMR run still uses a smaller
number of total and active cells than the ZEUS-3D run. We can save
noticeable computational time with SAGE/AMR as we tackle a problem
with more than a few million cells. This advantage will be bigger with
3D models.

{}
\clearpage

 \clearpage

 \clearpage

 \clearpage

 \clearpage

 \clearpage

 \clearpage

 \clearpage

 \clearpage

 \clearpage

 \clearpage

 \clearpage

 \clearpage

 \clearpage

 \clearpage

 \clearpage

 \clearpage

 \clearpage

 \clearpage

 \clearpage

 \clearpage

 \clearpage

\clearpage

\begin{table}
\begin{center}
\caption{Parameters for  
Starburst Models
\label{run} 
}
\begin{tabular}{lcccccc}
\\
   Model\tablenotemark{a}        & $\log_{10} \Sigma_0$\tablenotemark{b} & $\log_{10} L_{mech}$\tablenotemark{c} & 
$\dot{M_{in}}$\tablenotemark{d} & $R_{SN}$\tablenotemark{e} & $T_{floor}$\tablenotemark{f} & resolution \\

                & ($\mbox{ M}_{\odot}\mbox{ pc}^{-2}$) & (erg s$^{-1}$) & 
($\mbox{ M}_{\odot} yr^{-1}$) & (pc) & (K) & (pc) \\
\hline
X1    & 4 & 43 & 17  & 50 &  $10^4$ & 0.2     \\
X1-0  & 4 & 43 & 17  & 25\tablenotemark{g} &  $10^4$ & 0.1     \\ 
X1-2  & 4 & 43 & 17  & 50 &  $10^4$ & 0.4\tablenotemark{h}     \\ 
X1-4  & 4 & 43 & 17  & 50 &  $10^4$ & 0.8\tablenotemark{h}    \\ 
U1    & 4 & 43 & 17  & 50 & $10^2$ & 0.2     \\
U1-A   & 4 & 43 & 1.7 & 50 &  $10^2$ & 0.2     \\
U1-B   & 4 & 43 & 49  & 50 &  $10^2$ & 0.2     \\
U1-C   & 4 & 43 & 120 & 50 &  $10^2$ & 0.2     \\
X2    & 4 & 42 & 17  & 50 &  $10^2$ & 0.2     \\
X3    & 4 & 41 & 17  & 50 &  $10^2$ & 0.2     \\
S1    & 4 & 43 & 17  & 25 &  $10^2$ & 0.2    \\
V1
& 4.7 & 43 & 17 & 50 &  $10^2$ & 0.2     \\
\end{tabular}
\tablenotetext{a}{All models run with ZEUS}
\tablenotetext{b}{Central surface density}
\tablenotetext{c}{Mechanical luminosity}
\tablenotetext{d}{Mass loading rate}
\tablenotetext{e}{Size of source region where supernova energy is injected}
\tablenotetext{f}{The minimum temperature floor for cooling}
\tablenotetext{g}{The level of noise on the surface of the source region is kept same with that of our standard 
model, X1 by setting the number of cells covering the source region the same. }
\tablenotetext{h}{Ratioed grids are used (the resolution is 0.2pc within the source regions)}
\end{center}
\end{table}

\begin{table}
\begin{center}
\caption{Parameters for Dwarf Galaxy Models 
\label{sz}
}
\begin{tabular}{llcccccccc}
\\
model & code  & initial grid & res\tablenotemark{1} & cool & AMR\tablenotemark{2} & cycle & cells\tablenotemark{3} & cpu\tablenotemark{4} & procs\tablenotemark{5}  \\
& & & (pc) &  & &  & active (total) & per cycle &  \\
\hline
UN & SAGE &  $800\times1120$                   & 0.277  & OFF & OFF & 7244  & 896       & 4.0   & 12 \\
UC & SAGE &  $800\times1120$                   & 0.277  & ON  & OFF & 10877 & 896       & 4.9   & 12 \\
AN & SAGE &  $50\times70$                      & 0.277  & OFF & 5   & 7469  & 111 (146) & 0.86  & 12 \\
AC & SAGE &  $50\times70$                      & 0.277  & ON  & 5   & 10790 & 103 (136) & 0.84  & 12 \\
BC & SAGE &  $50\times70$                      & 0.139  & ON  & 6   & 22568 & 240 (319) & 1.5   & 24 \\
CC & SAGE &  $50\times70$                      & 0.0693 & ON  & 7   &
54000 &        503 (698)& 2.3   & 48 \\
RN & ZEUS &  $400\times878$\tablenotemark{6}   & 0.277  & OFF & OFF & 6035  & 351.2     & ?0.51 & 8 \\
RC & ZEUS &  $400\times878$\tablenotemark{6}   & 0.277  & ON  & OFF & 9036  & 351.2     & 3.7   & 8  \\ 
\end{tabular}
\tablecomments{We give the number of cycles, the average numbers of active (with AMR) and total cells, and 
the average cpu time spent per cycle that are used
to run the simulations until the bubbles blow out of the disk. 
}
\tablenotetext{1}{The highest resolution employed in the simulations.}
\tablenotetext{2}{The level of refinement used in AMR: if OFF, uniform or ratioed grids are used.}
\tablenotetext{3}{Average numbers of active and total cells used per
  cycle, in thousands.}
\tablenotetext{4}{The amount of cpu time spent per cycle.}
\tablenotetext{5}{Number of processors used for computation.}
\tablenotetext{6}{Ratioed grids are used.}
\end{center}
\end{table}

\end{document}